\documentclass[3p]{elsarticle}
\usepackage{bbding}
\usepackage{lineno,hyperref}
\usepackage[ruled,linesnumbered]{algorithm2e}
\usepackage[normalem]{ulem}
\usepackage{amsmath,amssymb,amsfonts}
\usepackage{graphicx}
\usepackage{textcomp}
\usepackage{xcolor}
\usepackage{epsfig, subfigure, amsmath, amssymb, wrapfig}
\usepackage{epstopdf}
\usepackage{balance}
\usepackage{color}
\usepackage{url}
\usepackage{pifont}
\usepackage{paralist}
\usepackage{commath}
\usepackage{enumitem}
\usepackage{balance}
\usepackage{comment}
\usepackage{xcolor}
\usepackage{soul}
\usepackage{graphicx}
\usepackage{float}
\usepackage{mwe}
\usepackage{caption}
\usepackage{subcaption} 
\usepackage{hyperref}
\usepackage{soul}
\usepackage{xcolor}
\usepackage{amsmath}
\usepackage{url}
\usepackage{epsfig, amsmath, amssymb}
\usepackage{multirow}
\usepackage{tikz}
\newcommand*\circled[1]{\tikz[baseline=(char.base)]{
            \node[shape=circle,fill=black, text=white, inner sep=1pt, font=\footnotesize] (char) {#1};}}

\SetKwRepeat{Do}{do}{while}
\definecolor{red}{rgb}{1,0,0}
\definecolor{green}{rgb}{0.0078 ,0.54,0.058}
\modulolinenumbers[5]
\makeatletter
\def\ps@pprintTitle{%
  \let\@oddhead\@empty
  \let\@evenhead\@empty
  \let\@oddfoot\@empty
  \let\@evenfoot\@oddfoot
}
\makeatother
\begin{document}

\begin{frontmatter}

% paper title
\title{SPARE: Securing Progressive Web Applications Against Unauthorized Replications}

% \author{Nur~Imtiazul~Haque\corref{cor1}\fnref{fn1}}
% \ead{email@uni.edu}
% \cortext[cor1]{Corresponding author}
% \fntext[fn1]{Student}

% \author{Author Two\fnref{fn2}}
% \ead{email2@uni.edu}
% \fntext[fn2]{Lecturer}

% \address{Address Here}

\author[]{Sajib~Talukder~\corref{cor1}\fnref{fn1}}
\author[]{Nur~Imtiazul~Haque~\fnref{fn2}}
\author[]{Khandakar~Ashrafi~Akbar~\fnref{fn3}}

\address[fn1]{George Washington University, USA}
\address[fn2]{University of Cincinnati, USA}
\address[fn3]{University of Texas at Dallas, USA}

\cortext[cor1]{Corresponding author}

\begin{abstract}
%

% WebView applications are widely used in mobile applications to display web content directly within the app, enhancing user engagement by eliminating the need to open an external browser and providing a seamless experience. Progressive Web Applications (PWAs) further improve usability by combining the accessibility of web apps with the speed, offline capabilities, and responsiveness of native applications. However, malicious developers can exploit this technology by duplicating PWA web links to create counterfeit native apps, thereby monetizing through user diversion. This unethical practice poses significant risks to both users and the original application developers, underscoring the need for robust security measures to prevent unauthorized replication. The issue of one way communication of Trusted Web Activity and Progressive Web Apps, we propose a practical solution to defend against or mitigate such attacks. We analyze the vulnerabilities of our proposed security solution to assess its effectiveness and introduce advanced measures to address any identified weaknesses, presenting a comprehensive defense framework. As part of our work, we developed a prototype web application that secures PWAs from replication by embedding a combination of Unix timestamps and device identifiers into the query parameters. We evaluate the effectiveness of this defense strategy by simulating an advanced attack scenario. Additionally, we created a realistic dataset reflecting mobile app user behavior, modeled using a Zipfian distribution, to validate our framework.

WebView applications are widely used in mobile applications to display web content directly within the app, enhancing user engagement by eliminating the need to open an external browser and providing a seamless experience. Progressive Web Applications (PWAs) further improve usability by combining the accessibility of web apps with the speed, offline capabilities, and responsiveness of native applications. However, malicious developers can exploit this technology by duplicating PWA web links to create counterfeit native apps, monetizing through user diversion. This unethical practice poses significant risks to users and the original application developers, underscoring the need for robust security measures to prevent unauthorized replication. Considering the one-way communication of Trusted Web Activity (a method for integrating web content into Android applications) and PWAs, we propose a query parameter-based practical security solution to defend against or mitigate such attacks. We analyze the vulnerabilities of our proposed security solution to assess its effectiveness and introduce advanced measures to address any identified weaknesses, presenting a comprehensive defense framework. As part of our work, we developed a prototype web application that secures PWAs from replication by embedding a combination of Unix timestamps and device identifiers into the query parameters. We evaluate the effectiveness of this defense strategy by simulating an advanced attack scenario. Additionally, we created a realistic dataset reflecting mobile app user behavior, modeled using a Zipfian distribution, to validate our framework.
\end{abstract}

\begin{keyword}
Progressive Web Apps, Trusted Web Activity, Machine learning, Cyber Security, Cyber Attacks, Threat Analysis, Formal Model.
\end{keyword}

\end{frontmatter}

\section{Introduction}

A Progressive Web Application (PWA) leverages web technologies to provide a user experience similar to a native application, making them particularly suitable for mobile devices due to their responsiveness and adaptability to various screen sizes. As the digital landscape continues to evolve, PWAs have emerged as a popular solution, combining the accessibility of web applications with the functionalities of native apps. PWAs offer enhanced user experiences through features like offline functionality, push notifications, and app-like interactions directly within a web browser~\cite{Chakradhar_OptimalStratigies}. This innovation, however, introduces unique security challenges that demand careful consideration to ensure the integrity, confidentiality, and availability of data and user interactions~\cite{Lee2018Pride}. Despite their many advantages, PWAs inherit and can amplify several security risks common to web applications, primarily due to their browser-centric architecture and reliance on third-party scripts~\cite{Carlo_D'Agnolo_2024_cside}. At their core, PWAs function as micro-web environments, running directly within the user's browser or a dedicated PWA window~\cite{How_phishing_PWAs_works_Kaspersky_2024}. By 2027, the PWA market is set to reach \$10.44 billion with a growth rate of about 32\%~\cite{emergen2021}. However, this popularity also brings increased scrutiny and potential security threats, particularly regarding the confidentiality of user information. Unauthorized replication extends beyond simple data duplication to encompass the illicit copying, deployment, or manipulation of an entire web application or its core functionalities by unauthorized entities~\cite{Medeiros_N_2015_duplicated}. This multifaceted threat presents complex challenges, as it can lead to data breaches, intellectual property theft, service disruption, and significant reputational and financial damage~\cite{Anton_Lohvynenko_2025}. While extensive efforts have been dedicated to mitigating various forms of web attacks, a comprehensive and systematized approach to securing web applications specifically from unauthorized replications remains a critical area of ongoing research and development~\cite{ServerSide_Xiaowei_Li_Yuan_Xue_2014}.
A malicious developer can execute a replay attack by creating and distributing a WebView-based application that embeds the URL of a target PWA, deceiving users by displaying the actual web app within the WebView. Adversaries may have several motivations for creating such applications. Firstly, the malicious app can inject its own advertisements into the WebView, earning direct monetary benefits from victims while harming the reputation of the legitimate organization and causing financial losses. Secondly, the malicious app can intercept user information and exploit it for fraudulent activities. Third, it can deploy hidden malicious payloads to victims' devices, using WebView as a facade for broader attacks, such as distributing ransomware.

Although replay attacks using malicious WebView applications pose significant threats, we are unaware of any research efforts to counter them. We propose a novel approach to secure PWAs by embedding dynamic query parameters into their URLs. By encrypting two key query parameters: Unix Time Frame and Device ID on the Android side using symmetric encryption, we can significantly enhance the security posture of our application. The Unix Time Frame serves multiple purposes; it provides a constantly changing value that limits the validity of any copied URL, thereby rendering it ineffective shortly after its creation. This time-sensitive mechanism ensures that even if an adversary gains access to a URL, it will become invalid after a predetermined period. The Device ID, however, acts as an additional layer of security. In scenarios where adversaries attempt to create dynamic review applications by leveraging the original app, Device ID allows our backend to recognize and limit access based on specific devices. This strategy mitigates the risk of unauthorized access, as each Device ID is unique and tied to a particular user device. By employing these two parameters, our approach not only fortifies the security of the PWA but also ensures that sensitive information remains protected from standard exploitation methods. This paper delves deeper into implementing this security mechanism, discusses its effectiveness against various attack vectors, and outlines the potential implications for developers and users in the PWA ecosystem. Through this exploration, we aim to contribute to the ongoing discourse on enhancing security in web applications, ultimately fostering a safer digital environment for all users. We name our proposed framework as \textbf{S}ecurity-Driven \textbf{P}rotection \textbf{A}gainst \textbf{R}eplication \textbf{E}xploits (SPARE). The principal contribution of our work is threefold. 
\vspace{-3pt}
\begin{itemize}
    \item We develop a PWA and instrument it with our proposed security features to protect against user diversion caused by malicious applications.
    \item We considered a realistic attack model and demonstrated a critical attack capable of evading the implemented security measures for a practical and effective security analysis.
    \item We strengthen the proposed security solution by enforcing a robust security policy, evaluating the vulnerability of our security analysis, and thus propose a defense guide.
\end{itemize}

All implementation and evaluation results are reproducible with the source codes on GitHub~\cite{securepwa2025, securepwaandroid2025, securepwabackend2025}. The rest of the paper is organized as follows: we provide an overview of the PWA, WebView applications, and other background information in Section~\ref{sec:background}. We formally describe the problem domain and the attack model in Section~\ref{sec:problem-formulation_attack-model}. In the following section, we present the technical details of the proposed SPARE framework (\ref{sec:technical-details}). We provide case studies to give insights about our proposed framework's working principle and capabilities in Section~\ref{sec:case-studies}. We evaluate SPARE using our synthesized realistic datasets based on Zipfian distribution in Section~\ref{sec:evaluation}. We conclude the paper in Section~\ref{sec:conclusion}.
\section{Background}
\label{sec:background}

% PWA is a web-based application that leverages standard web technologies to deliver an experience comparable to a native app designed for a specific platform. It works across various devices and platforms using a single codebase, much like a traditional website. However, it also offers features typical of native apps, such as being installable, functioning offline or in the background, and integrating with the device and other apps. PWAs bring together the best aspects of websites and native applications. PWAs have a standalone feature. It does not require any enforcement of any webview. 

A PWA is a web-based application that uses standard web technologies to provide an experience comparable to a native app built for a specific platform. It runs on various devices and platforms from a single codebase, much like a traditional website; however, it also incorporates native app–like features such as installability, offline or background functionality, and integration with the device and other apps. PWAs combine the best aspects of websites and native applications, and can run in standalone mode without requiring a WebView container.

%[https://developer.mozilla.org/en-US/docs/Web/Progressive_web_apps]. 
%[https://developer.mozilla.org/en-US/docs/Web/Progressive_web_apps/Guides/What_is_a_progressive_web_app]

% \begin{itemize}
% \item Trusted Web Activity
% \item Bubble web activity
% \item Security(Symmetric key encryption)
% \item Query parameter
% \item Service Worker
% \item Webmanifest
% \end{itemize}

% Vulnerabilities:
% symmetric encryption
% AES
% 1. https://link.springer.com/chapter/10.1007/978-3-030-75100-5_26
% 2. https://ieeexplore.ieee.org/abstract/document/9952818?casa_token=7sDTcYgnwkwAAAAA:rfXmcZVZK3xj7HEKm_V2-Sw0o5-rGqvETnoDAQVIgYD1Fkqyhbe6phvQP_HAVoC534EWDvJJ

\subsection{Background Study}
PWAs represent a revolutionary approach to web development, blending the universal accessibility of websites with the rich functionality traditionally associated with native mobile applications~\cite{biorn2017progressive}. Using modern web technologies such as service workers~\cite{developermozillaserviceworkers2025}, manifest files, and HTTPS protocols, PWAs deliver features that users typically expect only from installed applications: offline access, push notifications, and background data synchronization. This versatility has made PWAs increasingly attractive to developers seeking to create cross-platform experiences without navigating the complexities of app store deployment processes~\cite{fortunato2018progressive}. They are accessible via a unique URL and can be installed to a device's home screen \cite{Web_App_Features,PWA_as_Alternative}. PWAs aim to combine the best features of both web and native applications, offering a low-friction experience while providing an app-like feel\cite{PWA_Businesses,PWA_as_Alternative}. Yet, this very openness that makes PWAs so accessible across platforms also creates significant security vulnerabilities. Among the most concerning yet understudied threats is the unauthorized replication of PWAs through WebView-based applications. PWAs have emerged as a powerful alternative to traditional native mobile applications, offering a unified development approach that runs on web browsers while providing a rich user experience~\cite{fortunato2018progressive}. Google's concept of PWAs aims to normalize web development, allowing for centralized app creation that functions across various devices and platforms~\cite{fortunato2018progressive}. This capability, while beneficial for developers and users, simultaneously broadens the attack surface for malicious actors~\cite{DeView_Lee_2022}. In these scenarios, malicious actors can develop mobile applications that simply embed a WebView component displaying a legitimate PWA's URL. This effectively clones the original application's functionality and interface without requiring access to its source code. These unauthorized clones are then distributed as standalone mobile applications, often injected with advertisements, tracking code, or phishing mechanisms, directly undermining the original developer's intellectual property, revenue streams, and user trust.

PWAs have demonstrated increasing prevalence across digital platforms. Prominent social media entities, including Facebook, Instagram, X (formerly Twitter), and YouTube, have implemented PWA functionality. The integration of features such as application-like interfaces and cross-platform compatibility enables developers to engineer Android applications through the Trusted Web Activity (TWA)~\cite{developergooglepwainplay2025} methodology, subsequently facilitating distribution via the Play Store ecosystem. PWA-based Android applications can be integrated with Trusted Web Activities (TWAs) to allow for publishing on app stores like the Google Play Store\cite{PWA_TWA,deploying_pwa}. A PWA by itself requires a browser to run, but TWAs integrate the web experience with a native one, enabling an existing web app to function as a mobile application \cite{deploying_pwa}. This combination leverages the benefits of Progressive Web Apps (PWAs) along with the native-like features provided by Trusted Web Activities (TWAs)\cite{Web_App_Features, PWA_Businesses}.  PWAs operating within TWA architecture necessarily require browser infrastructure support, predominantly through Chrome Engine implementation. Although browsing histories are recorded within browser repositories, this presents a potential vulnerability wherein malicious actors may extract URLs from historical records. This vulnerability enables the unauthorized creation of WebView applications.

While WebView applications demonstrate limitations regarding offline functionality and native interface performance~\cite{developerandroidnativebridge2024}, primarily functioning as website containers within application frameworks, they can establish communication with frontend systems through JavaScript Bridge integration. Despite this capability representing a functional compromise, the extraction and replication of URLs remain viable for WebView application development. The protection of PWA-based Android Applications (TWAs) from unauthorized duplication presents significant challenges. Currently, there is a bidirectional communication protocol using PostMessage~\cite{webdev2020} which requires Chrome version 115.0.5790.13 or higher~\cite{developerchromepostmessage2025}. Which means it requires at least Android Version 8. This communication is session-based. It means that the ability to send and receive messages between your TWA (Android) and PWA (web) exists only during the active session of the TWA. A CustomTabsSession binds the Android app and the browser. If the Chrome process is killed, or the TWA is backgrounded for too long, the session can be lost. 

There is another communication method that operates unidirectionally (Android → Frontend), utilizing query parameters~\cite{developechromequeryparams2020}. While API implementation could potentially address this security concern, such solutions introduce server dependencies and potential performance degradation during PWA loading procedures, thereby compromising user experience and architectural simplicity. The fundamental objective of PWA development is a simplified implementation methodology, which would be compromised by such complexity. Query parameter implementation represents an established unidirectional communication methodology. Our research introduces an approach to differentiate between legitimate application requests and potentially malicious access attempts through encrypted query parameter implementation.

\subsection{Motivation}

The recent surge in low-code and no-code development platforms, combined with inconsistent app store enforcement policies, has dramatically lowered the technical barriers for creating WebView wrappers around existing PWAs. This trend raises several serious concerns:

\noindent\underline{\textit{Brand Impersonation:}} Bad actors can easily present themselves as official app providers, misleading users and potentially damaging the reputation of legitimate brands.
    
\noindent\underline{\textit{Monetization Theft:}} Unauthorized wrapper applications frequently incorporate advertisements or tracking mechanisms that divert potential revenue away from the rightful developers.
    
\noindent\underline{\textit{User Security Compromises:}} These wrapped PWAs may contain malicious code injections, phishing overlays, or invasive tracking tools that compromise end-user security and privacy.
    
\noindent\underline{\textit{Infrastructure Exploitation:}} Replicated applications continue to rely on the legitimate application's backend systems, increasing server loads and operational costs without contributing to intended business metrics.

Despite these significant threats, current research literature and developer resources offer surprisingly limited guidance on detecting, preventing, or mitigating such attacks specifically targeting PWAs. The absence of standardized defense mechanisms leaves PWA developers vulnerable and forces them into reactive rather than proactive security postures. This paper aims to address this critical gap by proposing a comprehensive, multi-layered defense strategy to shield PWAs from unauthorized WebView replication.

\subsection{Literature Review}

PWAs are web applications that behave like native apps, offering offline support, push notifications, and installation to the home screen. Trusted Web Activities (TWAs) extend this by allowing PWAs to run in full-screen mode within Android apps, using Chrome Custom Tabs under the hood. Although these technologies improve cross-platform development and user experience, they expose the application to risks not typically associated with native apps, primarily due to their reliance on web technologies and browser containers. Efforts to mitigate the security risks associated with PWAs and prevent unauthorized replication involve a combination of best practices and advanced security measures. A fundamental requirement for PWAs is the use of HTTPS, which encrypts all data transferred between the user's browser and the server, protecting sensitive information from interception and enabling critical PWA features like service workers and push notifications~\cite{best_practices_pwa,Web_App_Features,PWA_as_Alternative}. Although approaches like DeView have shown promise in eliminating accessible web APIs and preventing known exploits, the inherent flexibility and vastness of web APIs continue to present a significant attack surface~\cite{DeView_Lee_2022}. Researchers also highlight the need for continuous monitoring of security threats and ongoing solutions, especially since a large percentage of attacks occur at the application layer~\cite{Holistic_approch_web}.\\ 

Unauthorized replication of web applications, including PWAs, is often facilitated by a combination of inherent design complexities and common web vulnerabilities. Modern web applications thrive on delivering engaging experiences, but this quest for innovation can inadvertently introduce vulnerabilities~\cite{Debugging_PWA_Security_Issues_MoldStud_2025}. Statistics indicate that a significant percentage of cyberattacks target web applications, with nearly 75\% of organizations experiencing such attacks~\cite{Debugging_PWA_Security_Issues_MoldStud_2025}. The one-way communication using query parameters enables Trusted Web Activity(TWA) to communicate with the PWA. Due to browser history access in PWA apps, users can see the URL, making it easy to create WebView apps by simply changing the URL. PWAs can be rendered in mobile Web Views embedded within native mobile applications, which introduces security risks at the intersection of application and web security~\cite{web_applications_mobile_platforms}.
Attackers can display inappropriate ads on top of the WebView to any age group, as their primary goal is to copy the main app to generate revenue. For financial transaction applications like ticket-buying websites, attackers can copy the PWA app and redirect users to malicious transaction sites from the top layer.

Digital Asset Links (DAL)~\cite{developergoogledigitalassets2024} add a layer of protection against attackers creating TWA apps. Without valid asset links, the address bar will remain visible at the top of attacker-made TWA apps. When an attacker's TWA doesn't match the fingerprint (SHA-256), the app launches with the address bar displayed. The developers of the Trusted Web Activity recommended using DAL to verify domain ownership in TWA.
For two-way communication, JavaScript User Interface (JSUI)~\cite{fortunato2018progressive} can work, but it presents security issues and requires manual caching of HTML/JS for offline support. Performance-wise, PWA is significantly better than WebView. Communication from a Progressive Web App frontend to its Android wrapper via Trusted Web Activity presents several challenges. While WebView supports bidirectional communication between JavaScript and native code, TWA lacks this capability. Despite TWA making it easy to bring PWAs to the Play Store, communication between the PWA and Android native shell remains restricted.

\subsubsection{WebView Wrapping and Security Risks}
WebView is a crucial component on both Android and iOS platforms, allowing mobile applications to embed a simplified yet powerful browser~\cite{Luo_Hao_Webview_Attacks_2011, Geetha_2017}. This enables apps to function as customized browsers for their intended web applications, with features that allow app code to interact with JavaScript within web pages, intercept events, and modify them via various APIs~\cite{Luo_Hao_Webview_Attacks_2011, Geetha_2017}. Progressive Web Applications (PWAs) can be rendered not only in standalone browsers like Google Chrome or Firefox but also within mobile WebViews embedded in native mobile applications~\cite{web_applications_mobile_platforms}. Securing PWAs from unauthorized replication is a critical area of concern, given their increasing adoption and the native app-like features they offer~\cite{article_pwa, Lee2018Pride}. While much of the existing literature focuses on general PWA security vulnerabilities like those related to service workers and phishing, there's a distinct need to address the specific challenges posed by unauthorized replication~\cite{Lee2018Pride, ServiceWorkerSurver_Jeong_Junbeom_Hur_2022}. Many Android apps, including popular ones like Facebook, Twitter, and Instagram, utilize WebView to display web content without redirecting users to external browsers~\cite{Anuradha_Singh_Navneet_Goyal_2020, Imamura_Hiroyuki_2018}.

For PWAs specifically, this vulnerability is amplified due to their publicly accessible URL architecture, highlighting that while PWAs are typically designed with a domain-centric approach, they fundamentally lack binding mechanisms that would restrict their execution to trusted environments or containers. This architectural limitation makes them inherently susceptible to unauthorized embedding or misuse through WebView wrappers. The battle against unauthorized replication and other PWA security threats is an ongoing process that requires continuous learning, adaptation to new threats, and the implementation of smart, layered security solutions~\cite{Debugging_PWA_Security_Issues_MoldStud_2025}.

\subsubsection{Existing Client-Side and Server-Side Defenses}

Client-side defenses for Progressive Web Applications (PWAs) focus on securing the application directly on the user's device, leveraging browser features and application design to mitigate risks~\cite{Lee2018Pride}. These defenses are crucial given that PWAs provide native app-like browsing experiences and utilize advanced HTML5 features such as service workers and caching~\cite{Lee2018Pride}. Common client-side defenses such as Content Security Policy (CSP) and \texttt{X-Frame-Options} headers are frequently recommended to prevent framing attacks and unauthorized embedding. However, these protective measures were primarily designed to prevent iframe inclusion on websites and offer limited protection against mobile WebView-based embedding, particularly when WebView clients deliberately override or ignore these security policies~\cite{developermozilla2025}. 

Server-side defenses are essential for securing PWAs by protecting the backend infrastructure, data, and communication channels~\cite{Mohsen_Haadi_Jafaarian_2019}. These measures aim to counteract attacks that originate from or target the server, complementing client-side protections. Server-side protection techniques have also been explored, including User-Agent filtering, Origin and Referer header validation, and token-based access control systems. While these approaches can help identify unauthorized access patterns, sophisticated attackers can easily circumvent them by spoofing request headers. More robust solutions like Google's Play Integrity API~\cite{google_integrity} and Apple's DeviceCheck~\cite{apple_device_check,apple_integrity} provide attestation mechanisms that allow native applications to verify the authenticity of client devices or applications making requests. Unfortunately, these powerful mechanisms aren't inherently compatible with the PWA architecture and require native integration components, effectively leaving standalone PWAs without access to such sophisticated attestation layers. While PWAs and TWAs offer performance and UX benefits, their openness to WebView-based duplication remains a critical vulnerability.
\section{Problem Definition and Attack Modeling}
\label{sec:problem-formulation_attack-model}

This section provides a formal definition of the PWA replication attack and the attack model considered in the SPARE framework.

\subsection{Problem Definition}
Suppose a PWA $\mathbb{P}$ provides services to a set of clients $\mathcal{C}$ through a weblink $\mathbb{W}$. An adversary $\mathbb{A}$ has developed a malicious WebApp, denoted as $\mathbb{P}^A$, which replicates the functionality of $\mathbb{P}$. The malicious WebApp redirects users of $\mathbb{P}^A$ to $\mathbb{W}$, allowing the adversary to gain financial benefits or access confidential information.

A client $c \in \mathcal{C}$ is considered legitimate if they access $\mathbb{W}$ via $\mathbb{P}$, and malicious if they access $\mathbb{W}$ through $\mathbb{P}^A$. To prevent unauthorized access to $\mathbb{W}$ from applications other than $\mathbb{P}$, we propose embedding fingerprinting information in the form of query parameters in the URL of $\mathbb{W}$.

Initially, the encrypted UTC time of the client's device, $\bar{t}$, is used as the sole query parameter. A client $c$ can access resources from a dedicated server $\mathbb{S}$ using a URL of the form $w \, ? \, \bar{t}$, where $w \in \mathbb{W}$, $c \in \mathcal{C}$, and $\bar{t} = \text{enc}(\text{deviceTime}(c))$. Here:
\begin{itemize}
    \item $\text{enc}(.)$ encrypts data using a symmetric encryption key shared between $\mathbb{P}$ and $\mathbb{S}$.
    \item $\text{deviceTime}(c)$ returns the UTC time of the client $c$'s device.
\end{itemize}

The server $\mathbb{S}$ denies access if it receives a request for $w$ without any query parameters. For a URL with valid syntax, $w \, ? \, \bar{t}$, $\mathbb{S}$ decrypts $\bar{t}$ to obtain the client’s UTC time, $t = \text{dec}(\bar{t})$. If the difference between $t$ and the server's current UTC time, $t_s$, exceeds a predefined threshold $\mathrm{T}$ (set to 1 minute in our experiment), the server denies the request. This mechanism reduces the attack surface since static weblinks $\mathbb{W}$ used by adversaries in $\mathbb{P}^A$ are ineffective without appropriate query parameters.

However, an adversary could access a valid URL ($w_a \in \mathbb{W}$, with query parameters $\bar{t_a}$) by interacting with $\mathbb{P}$ through their browser, as explained in Section~\ref{sec:background}. Despite this, such URLs would become invalid since by the time a malicious client accesses $\mathbb{W}$ through $\mathbb{P}^A$, the server's UTC time $t_s$ would significantly differ from $t_a = \text{dec}(\bar{t_a})$ (i.e., $t_s - t_a \gg \mathrm{T}$). In our work, we consider a more sophisticated adversary. This adversary maintains a database $\mathbb{D}$ that is updated every $\Delta t$ (where $\Delta t < \mathrm{T}$) by retrieving a valid URL from $\mathbb{P}$ through their browser. When a client opens the malicious WebApp $\mathbb{P}^A$, the WebView in $\mathbb{P}^A$ loads the most recently updated link from $\mathbb{D}$ instead of a static URL, effectively bypassing the defense. Consequently, $\mathbb{S}$ cannot differentiate between malicious traffic and legitimate traffic. Therefore, developing an effective strategy to mitigate such attacks is crucial.

\subsection{Attack Model}
\label{subsec:attack-model}
An attack model is a structured framework that defines an adversary's assumptions, capabilities, and goals in a security scenario. It is used to analyze and simulate potential threats to a system systematically. Here, we provide the characteristics of the attack and the adversary, focusing on the following components.
%%%%%%%%%%%%%%%%%%%%%%%%%%%%%%%%%%%%%%%%%%%%%

%%%%%%%%%%%%%%%%%%%%%%%%%%%%%%%%%%%%%%%%%%%%%
%
\subsubsection{Attack Technique}
When implementing authentication for PWAs via query parameters, developers face significant security challenges due to browser history accessibility. This analysis examines potential attack vectors against query parameter-based authentication systems.\\

\noindent The adversary creates a malicious WebApp that mimics the legitimate service provider. This malicious app redirects users to official weblinks while allowing the adversary to gain financial benefits or access sensitive data. To counter this, the legitimate provider embeds encrypted timestamps in the URLs to validate access requests. The server verifies whether the timestamp falls within an acceptable time window, blocking outdated or missing timestamps to prevent unauthorized access. The core vulnerability is due to PWAs in TWAs leaving traces in browser history, query parameters used for authentication being visible in these URL histories, and Webview applications loading these URLs with their authentication parameters.

\noindent A more advanced adversary maintains an updated database of the valid URL by frequently extracting fresh timestamps from the legitimate app through sniffing on the browser cache. This database is updated at intervals shorter than the allowed time window for validation, ensuring the adversary always serves recent, valid links to users. As a result, the server cannot distinguish between legitimate and malicious access, effectively bypassing the defense mechanism. The step-by-step attack process is described as follows.

\noindent\underline{\textit{Initial Access:}} An attacker first obtains the legitimate application and installs it on one or more devices. This provides access to the structure of the authentication query parameters, the pattern and frequency of parameter updates, and any observable encryption or encoding patterns.

\noindent\underline{\textit{History Extraction:}} The attacker develops a system to extract browser history from devices running the authentic application. This involves creating a background service that monitors Chrome/browser history, focusing specifically on URLs matching the target PWA's domain, and extracting full URLs including authentication query parameters.

\noindent\underline{\textit{URL's Cloud Infrastructure:}} A cloud-based collection system is established. Devices with the authentic app installed upload extracted URLs to a centralized database. The system tags each URL with metadata, including a timestamp and the source device. These URLs are then indexed and stored for distribution.

\noindent\underline{\textit{Dynamic Webview:}} The attacker creates a webview-based clone application that connects to the cloud repository upon launch, requests a fresh, unused authentication parameter, dynamically constructs and loads the URL with this parameter, and reports back when the parameter has been used.

\noindent\underline{\textit{Mark URL:}} To avoid detection through duplicate usage patterns, each authentication parameter is marked as "used" after distribution. The system prioritizes the newest parameters to maximize validity periods and implements load balancing across multiple harvesting devices.

\noindent\underline{\textit{Scaling the Attack:}} To overcome defensive measures like time-based thresholds, multiple harvesting devices are deployed to generate a continuous stream of valid parameters. The attack infrastructure may distribute different harvesting devices across various IP addresses and locations. Each harvesting device maintains normal usage patterns to avoid triggering anomaly detection.

\noindent\underline{\textit{Threshold Circumvention:}} 
When faced with per-device thresholds, the attacker expands their device pool, potentially using emulated devices. Parameters are harvested from multiple devices to stay under individual threshold limits. The system may also incorporate device rotation strategies to distribute harvesting activities.

\subsubsection{\textit{Attack Goal}}
The primary goal of the attack is to bypass the server's security measures and ensure that users of the malicious WebView app can access the weblinks without interruption. Suppose the most recently updated weblink in the database is $w \, ? \, \bar{t_{a'}}$, where $t_{a'} = \text{dec}(\bar{t_{a'}})$. The attack will succeed only if the time difference between the server's current time $t_s$ and $t_{a'}$ satisfies $t_s - t_{a'} < \mathrm{T}$; otherwise, the malicious clients will be denied access.

\subsubsection{Attack Assumptions}
We consider the following assumptions while formalizing our attack model.

\noindent\underline{\textit{Assumption I:}} The adversary continuously interacts with the legitimate PWA application to obtain the latest valid URL. The firewall of the legitimate server application cannot detect and block the adversary. Adversaries may employ various spoofing techniques to evade detection.

\noindent\underline{\textit{Assumption II:}} The adversary-created WebView application can dynamically load content by fetching the URL from the adversary's database.

\noindent\underline{\textit{Assumption III:}} The adversary is aware of the user validation process, including analyzing query parameters and the fact that the query parameter encrypts the timestamp when a client initiates a request. However, the encryption key remains unknown to the adversary.  

\noindent\underline{\textit{Assumption IV:}} The adversary's database has sufficient capacity to process and handle requests from all malicious app clients attempting to retrieve a valid weblink.
%%%%%%%%%%%%%%%%%%%%%%%%%%%%%%%%%%
%%%%%%%%%%%%%%%%%%%%%%%%%%%%%%%%%%
%
%%%%%%%%%%%%%%%%%%%%%%%%%%%%%%%%%%%%%%%%%%%%%
\subsubsection{Adversarial Attributes}
The adversarial attributes encompass knowledge, capability, and accessibility. While we have already outlined these attributes in the assumptions, we categorize them as follows.

\noindent\underline{\textit{Knowledge:}} The adversary knows the encryption mechanism and that the query parameter contains an encrypted timestamp; however, they do not possess the encryption key. This characterizes the attack as a gray-box attack.

\noindent\underline{\textit{Accessibility:}} The adversary retrieves valid weblinks by extracting them from the adversary's copy of the legitimate application's network traffic or browser cache. The adversary can obtain the legitimate weblink containing the encrypted query parameters by sniffing on the browser cache or intercepting the application's communications.

\noindent\underline{\textit{Resources:}} The adversary maintains a database and possesses a copy of the malicious app. The database does not need to store all previously retrieved weblinks; instead, it only retains the most recently updated weblink at any given time. However, sufficient bandwidth is required to ensure all malicious app users can access the database content simultaneously.
%%%%%%%%%%%%%%%%%%%%%%%%%%%%%%%%%%%%%%%%%%%%%%%%%%%%%%%%%%%%%%%%%%%
%%%%%%%%%%%%%%%%%%%%%%%%%%%%%%%%%%%%%%%%%%%%%%%%%%%%%%%%%%%%%%%%%%%
%%%%%%%%%%%%%%%%%%%%%%%%%%%%%%%%%%%%%%%%%%%%%%%%%%%%%%%%%%%%%%%%%%%
\section{Technical Details}
\label{sec:technical-details}
In this section, we present the technical details of the SPARE framework. We begin by outlining the problem scope and proposing an initial defense strategy (see~\ref{subsec:problem-scope}). Next, we identify attack vectors that bypass this initial defense strategy (see~\ref{subsec:exploring-initial-defense}). We then enhance the defense strategy to develop SPARE, which is designed to counter the identified attack vectors (see~\ref{subsec:advancing-defense}). Finally, we explore an advanced attack technique aimed at bypassing SPARE and assess its effectiveness (see~\ref{subsec:advanced-attack}).

%%%%%%%%%%%%%%%%%%%%%%%%%%%%%%%%%%%%%%%%%%%%%%
\begin{wrapfigure}{r}{0.45\columnwidth}
\includegraphics[width=0.45\columnwidth]{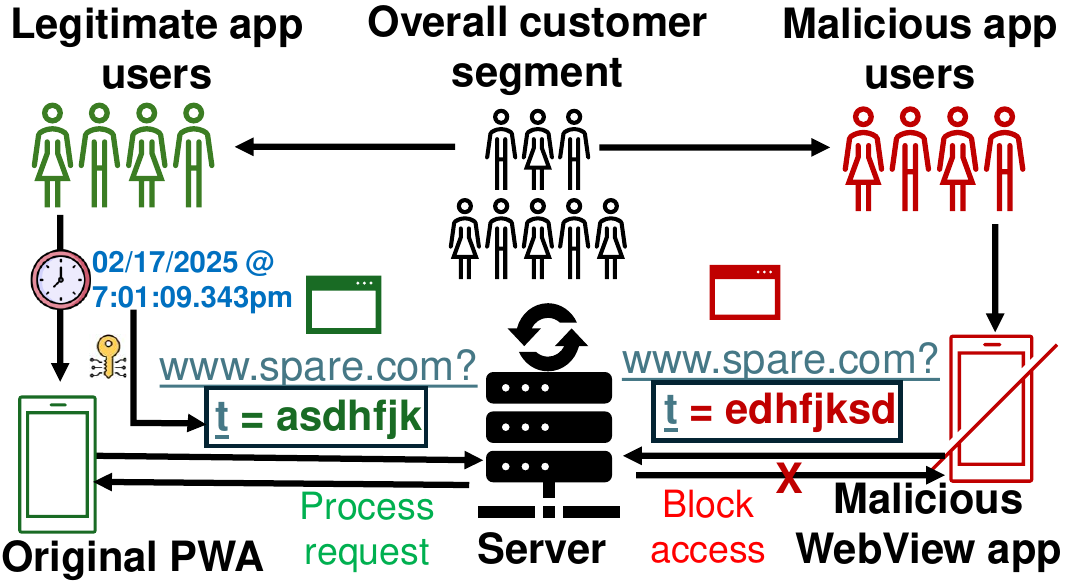}
\caption{Overview of the problem scope and initial defense strategy.}
\label{fig:problem-overview}
\vspace{-10pt}
\end{wrapfigure}
%%%%%%%%%%%%%%%%%%%%%%%%%%%%%%%%%%%%%%%%%%%%%%

%%%%%%%%%%%%%%%%%%%%%%%%%%%%%%%%%%%%%%%%%%%%%%%%%%
% \begin{figure}[t]
% \centering
% \includegraphics[width = 0.5\columnwidth]{figures/SPARE-Problem-Overview.pdf}
% \caption{Overview of the problem scope and initial defense strategy.}
% \label{fig:problem-overview}
% \end{figure}
% %%%%%%%%%%%%%%%%%%%%%%%%%%%%%%%%%%%%%%%%%%%%%%%%%%

\subsection{Problem Scope and Initial Defense Strategy}
\label{subsec:problem-scope}
Figure~\ref{fig:problem-overview} illustrates the overall customer segmentation, divided into two groups: one using the legitimate application and the other redirected to a malicious application that closely resembles the original. The malicious mobile application utilizes a WebView to load content using the same URL as the original PWA. The adversary or malicious app developer can easily obtain the original weblink used by the PWA by analyzing network packets or inspecting browser history while interacting with the original PWA as a regular user. Consequently, the malicious application acts as a man-in-the-middle between the customer and the server, enabling the adversary to integrate additional features into the application to achieve their attack objectives. To mitigate such an attack, we propose a primary defense strategy that leverages query parameters in the weblink. Instead of allowing unrestricted access to the server, we introduce an application firewall that filters incoming traffic based on the content of the query parameter. The security enhancement strategy is depicted in Figure~\ref{fig:problem-overview}, where the PWA retrieves the current UTC time from the user's device, encrypts it using a symmetric encryption technique such as AES\cite{nistaes2001}, and embeds the resulting ciphertext containing the encrypted timestamp in the query parameter.

\begin{wrapfigure}{r}{0.35\columnwidth}
\includegraphics[width=0.35\columnwidth]{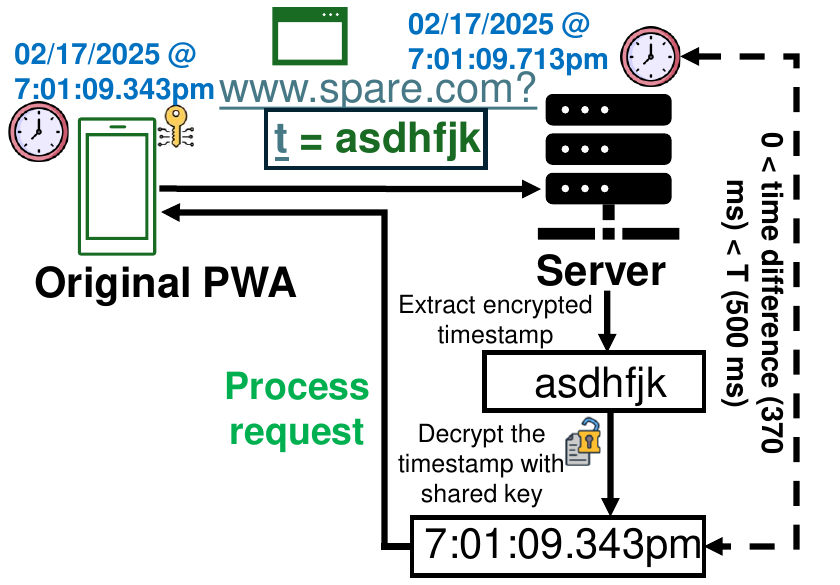}
\caption{Overview of the server operation.}
\label{fig:server-operations}
\vspace{-15pt}
\end{wrapfigure}

The server validation process is further illustrated in Figure~\ref{fig:server-operations}, where the server analyzes the query parameter before processing client requests. Since the server possesses the decryption key, it can decrypt the query parameter to extract the timestamp indicating when the request was initiated. After retrieving the timestamp, the server compares it with its UTC time. While there may be slight differences due to communication delays, even for legitimate requests, this delay remains within an acceptable threshold, assumed to be 1 hour in our problem scenario (as shown in Figure~\ref{fig:server-operations}). If the time difference is within the allowed threshold, the server considers the request legitimate, processes it, and responds to the client. However, in Figure~\ref{fig:problem-overview}, the malicious app user is denied access because, by the time they attempt to reuse a copied URL, the time difference far exceeds 1 hour, making the request invalid. 
%
%%%%%%%%%%%%%%%%%%%%%%%%%%%%%%%%%%%%%%%%%%%%%%
%
%
%%%%%%%%%%%%%%%%%%%%%%%%%%%%%%%%%%%%%%%%%%%%%%

% %%%%%%%%%%%%%%%%%%%%%%%%%%%%%%%%%%%%%%%%%%%%%%%%%%
% \begin{figure}[!htb]
% \centering
% \includegraphics[width = 0.7\columnwidth]{figures/Server-Operations-Alt.pdf}
% \caption{Overview of the server operations in the initial defense strategy.}
% \label{fig:server-operations}
% \end{figure}
% %%%%%%%%%%%%%%%%%%%%%%%%%%%%%%%%%%%%%%%%%%%%%%%%%

\subsection{Exploiting the Initial Defense Strategy}
\label{subsec:exploring-initial-defense}
\begin{wrapfigure}{r}{0.24\columnwidth}
\includegraphics[width=0.24\columnwidth]{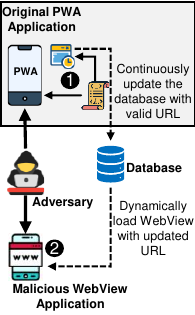}
\caption{Demonstration of the exploitation of the initial defense strategy.}
\label{fig:exploiting-initial-defense}
\vspace{-15pt}
\end{wrapfigure}

As we explore potential defense strategies to prevent PWA replication attacks, we aim to identify attack vectors capable of penetrating the system to assess the robustness of the proposed defense. The attack model for this exploit is detailed in Section~\ref{subsec:attack-model}. Figure~\ref{fig:exploiting-initial-defense} illustrates an adversary's steps to evade the proposed initial defense strategy. In this scenario, we consider an adversary who develops a malicious WebApp and maintains a copy of the original PWA application. The adversary executes \circled{1} an automated script that continuously interacts with the original PWA, extracting the most recent valid URL from the browser history on their device. This updated URL is then stored in a database. Any client using the malicious application will have their \circled{2} WebView dynamically populated with the URL retrieved from the database rather than a static URL.
%%%%%%%%%%%%%%%%%%%%%%%%%%%%%%%%%%%%%%%%%%%%%%
%%%%%%%%%%%%%%%%%%%%%%%%%%%%%%%%%%%%%%%%%%%%%%

The validity of the URL is determined by the server running the firewall, which assesses whether the URL was generated within an acceptable timeframe. The success of this intrusion hinges on two factors: the server threshold $\mathrm{T}$ and the speed at which the database updates by interacting with the original PWA. If the adversary's system can retrieve and distribute updated URLs quickly enough, the attack may bypass the server's validation mechanism, thus compromising the effectiveness of the initial defense strategy.

\subsection{Advancing the Defense}
\label{subsec:advancing-defense}
We assume the attacker can successfully evade the initial defense strategy, as discussed in Section~\ref{subsec:exploring-initial-defense}. To strengthen the defense, the SPARE framework introduces an enhanced mechanism that incorporates the device ID into the query parameter from the PWA application instead of solely embedding an encrypted timestamp. To further reinforce security, we introduce a new threshold, $\mathrm{R}$, designed to restrict an infeasible number of requests from the same client. This threshold is determined based on the activity patterns of the most active legitimate users, ensuring that normal usage remains unaffected while preventing excessive, automated, or replicated requests from a single source. Hence, with this enhanced defense mechanism, a client $c$  can access resources from the dedicated server $\mathbb{S}$ using a URL of the form $w ? \overline{td}$, where $w \in \mathbb{W}$, $c \in \mathcal{C}$, and $\overline{td} = \text{enc}(\text{deviceTime(c) $||$ deviceID(c)}$. Here, deviceTime(c) and deviceID(c) represent the timestamp and device ID of the client's device, respectively. We use the symbol ($||$) to denote concatenation.

The attack strategy that successfully bypasses the initial defense will fail against this advanced defense mechanism since all malicious app users rely on URLs generated from a single compromised device. As a result, the server will flag and block repeated requests originating from the same device ID once they exceed the defined threshold $\mathrm{R}$. The attack strategy that successfully bypasses the initial defense will fail to evade this advanced defense mechanism since all malicious app users rely on URLs generated from a single compromised device. As a result, the server will flag and block repeated requests originating from the same device ID once they exceed the defined threshold $\mathrm{R}$ in a single day. This enhanced security measure significantly limits the effectiveness of the defense, as the adversary can no longer distribute valid URLs among multiple users without being detected. 

To further strengthen the defense, we introduce an additional firewall rule that detects simultaneous requests from the same device. Such behavior indicates potential replication, a key characteristic of the attack. The system will flag it as suspicious activity if multiple requests are detected from the same device ID within a short window. We propose blocking access for device IDs that exhibit repeated, simultaneous request patterns, effectively neutralizing the attack by preventing the mass distribution of valid URLs. 

Our defense methodology implements a dynamic approach to secure application access through query parameters. Upon initial URL launch, a unique query parameter is generated and transmitted to the frontend. The frontend verifies this parameter and forwards it to the backend for authentication purposes.
A limitation exists where this query parameter is only generated upon the first launch. When the application is not running in the background, the system prompts users to close and reopen the application, creating a suboptimal user experience.

To address this limitation, we store the query parameter locally when the user launches the application. The request payload includes a "resubmit" parameter, which is set to true when a user is already in the application and initiating a server request. This eliminates the need for users to repeatedly close and reopen the application.
Server-side handling of the resubmit value requires careful implementation, as the query parameter remains unchanged, potentially accessing existing data. Database updates include refreshing the user's last request timestamp with each interaction.
The system maintains a rolling database architecture that compares current request times against previous request timestamps. Each request increments an access counter associated with the device ID. When a user reaches the defined threshold, they receive a notification to pause requests for a predetermined period, after which their access count resets, allowing them to regain access privileges.

\subsection{Advanced Attack Technique}
\label{subsec:advanced-attack}

\begin{wrapfigure}{r}{0.45\columnwidth}
\includegraphics[width=0.45\columnwidth]{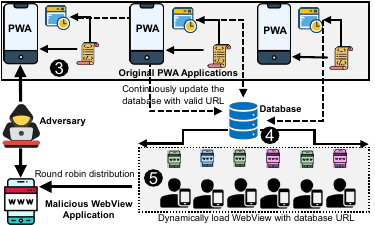}
\caption{Demonstration of the advanced exploit.}
\label{fig:advanced-exploit}
\vspace{-25pt}
\end{wrapfigure}

To thoroughly evaluate the robustness of our enhanced defense mechanism, we now consider an advanced attack strategy aimed at circumventing the newly introduced security measures. Figure~\ref{fig:advanced-exploit} illustrates this proposed exploit. In the previous attack scenario, the adversary could not spoof (i.e., modify) the device ID while storing updated URLs from the original PWA, as the URL generation process involved encryption. Consequently, distributing a single valid URL across multiple malicious app users is no longer viable, as exceeding the threshold $\mathrm{R}$ requests from the same device ID would be flagged as an anomaly.

%%%%%%%%%%%%%%%%%%%%%%%%%%%%%%%%%%%%%%%%%%%%%%
%%%%%%%%%%%%%%%%%%%%%%%%%%%%%%%%%%%%%%%%%%%%%%

To overcome this restriction, the attacker adapts their strategy as follows:  
\begin{enumerate}
    \item The adversary interacts with \circled{3} multiple copies of the original PWA from different physical devices. Each device independently retrieves valid URLs in real time.
    \item These URLs are then continuously updated in the \circled{4} centralized database accessible by the malicious applications.
    \item Instead of distributing URLs randomly among malicious app users, the attacker employs a \circled{5} round-robin distribution strategy. This approach systematically cycles through available URLs, ensuring that no single device ID exceeds the daily request threshold $\mathrm{R}$.
\end{enumerate}

By leveraging multiple devices and distributing URLs efficiently, the adversary reduces the likelihood of detection and denial from the server due to excessive request activity from a single device. This method represents a more sophisticated and scalable approach to bypassing the SPARE defense mechanism.

% \begin{figure}[h]
% \label{fig:system}
% \includegraphics[width=7cm,height=10cm]{figures/pwa.png}
% \caption{System Architecture}
% \end{figure}

% %%%%%%%%%%%%%%%%%%%%%%%%%%%%%%%%%%%%%%%%%%%%%%%%%%
% \begin{figure}[t]
% \centering
% \includegraphics[width = 0.65\columnwidth]{figures/SPARE_Advanced-Exploit.pdf}
% \caption{Demonstration of the advanced exploit.}
% \label{fig:advanced-exploit}
% \end{figure}
% %%%%%%%%%%%%%%%%%%%%%%%%%%%%%%%%%%%%%%%%%%%%%%%%%%
\section{Case Studies}
\label{sec:case-studies}
In this section, we present empirical case studies to demonstrate how the SPARE framework operates in analyzing the security of a PWA application. We begin with a benign scenario in Case Study 1, followed by subsequent case studies that showcase how the proposed defense strategy performs against attackers with varying levels of knowledge and skill. To enhance clarity and explainability, we use simplified numerical values in the case studies. The original data is used in the following section (Section~\ref{sec:evaluation}) for the formal evaluation of the SPARE framework.

%%%%%%%%%%%%%%%%%%%%%%%%%%%%%%%%%%%%%%%%%%%%%%%%%%%
%%%%%%%%%%%%%%%%%%%%%%%%%%%%%%%%%%%%%%%%%%%%%%%%%%%
\subsection{Case Study 1: Benign Scenario}
Here, we present a numerical study of a benign, attack-free scenario. We assume that 20 users actively use the PWA, each generating between 25 and 45 daily requests. To manage bandwidth usage and mitigate the risk of denial-of-service attacks, we introduce a threshold to limit excessive request rates. Table~\ref{tab:benign-threshold} presents the percentage of successful and failed requests at various per-day device access thresholds for authenticated users. The table shows that as the daily threshold increases, the number of successful requests (requests processed by the server) also rises. When the threshold is set to 45, the success rate reaches 100\%, as no user exceeds 45 requests per day in this scenario.

%%%%%%%%%%%%%%%%%%%%%%%%%%%%%%%%%%%%%%%%%%%%%%%%%%
\begin{table}[!t]
\scriptsize
\caption{Case Study of a Benign Scenario Demonstrating Percentage of Successful/Failed Requests with Varying Per-Day Device Access Threshold for an Authenticated User}
\label{tab:benign-threshold}
\begin{tabular}{|p{1cm}|p{1.8cm}|p{1.8cm}|p{1.5cm}|p{2.cm}|p{1.1cm}|p{2cm}|p{2cm}|}
\hline
Number of PWA Users & Minimum Number of Requests/User-Day & Maximum Number of Requests/User-Day & Total Requests/Day   & Average Number of Requests/User-Day & Threshold & Number of Successful Requests (Percentage of Successful Requests) & Number of Failed Requests (Percentage of Failed Requests) \\ \hline
\multirow{4}{*}{20} & \multirow{4}{*}{25}                 & \multirow{4}{*}{45}                 & \multirow{4}{*}{666} & \multirow{4}{*}{33.3}              & 30        & 600 (90\%)                                                        & 66 (10\%)                                                \\ \cline{6-8} 
                    &                                     &                                     &                      &                                     & 35        & 664 (99.7\%)                                                      & 2 (0.3\%)                                              \\ \cline{6-8} 
                    &                                     &                                     &                      &                                     & 40        & 666(100\%)                                                       & 0 (0.0\%)                                              \\ \cline{6-8} 
                    &                                     &                                     &                      &                                     & 45        & 666 (100\%)                                                       & 0 (0\%)                                                   \\ \hline
\end{tabular}
\end{table}

%%%%%%%%%%%%%%%%%%%%%%%%%%%%%%%%%%%%%%%%%%%%%%%%%%%
%%%%%%%%%%%%%%%%%%%%%%%%%%%%%%%%%%%%%%%%%%%%%%%%%%%
\subsection{Case Study 2: Amateur Attacker}
An amateur attacker is unaware of the complete encryption process. The attacker attempts to create a WebView application by simply copying the URL used in the original application, i.e., \texttt{http://spare.com}, and distributing the malicious app. However, users of this malicious application will be denied access, as the requests sent to the server will lack the required valid query parameters.

%%%%%%%%%%%%%%%%%%%%%%%%%%%%%%%%%%%%%%%%%%%%%%%%%%%
%%%%%%%%%%%%%%%%%%%%%%%%%%%%%%%%%%%%%%%%%%%%%%%%%%%
\subsection{Case Study 3: Naive Attacker}
A naive attacker lacks understanding of the internal mechanisms of the system. Specifically, they are unaware of the encryption-based validation used in the URL query parameters. However, unlike the amateur attacker, the naive attacker recognizes the importance of query parameters for accessing resources from the server. Suppose the attacker creates a malicious WebApp that visually mimics the original PWA and distributes it to a group of users. The attacker hardcodes a URL such as \texttt{http://spare.com?id=hrdqiuayf378\\ary389qhqueaid78} into the malicious application. The attacker may initially access the URL from their own browser using the original application. However, once the URL has been used, any subsequent attempts to access it—such as from the distributed malicious apps—will fail. This is because the server tracks previously used URLs for resource requests. Each query parameter includes an embedded device ID and timestamp indicating the request's initiation. If the same link is reused, the server detects that a request from the same device ID and timestamp has already occurred, signaling a replication attempt. As a result, the WebView in the malicious application fails to load the intended content, rendering the attack ineffective.

We now consider a scenario where the attacker does not connect the original application to the internet, preventing any requests from reaching the server. However, since the query parameter generation occurs on the front-end of the original PWA, a valid URL can still be generated locally and copied directly from the browser, even without an internet connection. As a result, the attacker can obtain a valid URL for distribution. This URL remains valid for a threshold duration, denoted as $\mathrm{T}$, which we assume to be 1 hour in this case. Suppose the attacker distributes the malicious WebApp, including the generated URL, to a user within this 1-hour window. In that case, the user can access the application for that day, provided they do not exceed the per-day device access threshold (e.g., 35 requests). However, the server will detect the replication attempt if multiple users attempt to access the application using the same URL. All requests from the second and subsequent users will be denied in such cases. Therefore, through this method, the naive attacker can at most enable a single user to access the malicious WebApp for one day. The session will expire on subsequent days, and any attempt to reuse the same URL will result in denied requests.

%%%%%%%%%%%%%%%%%%%%%%%%%%%%%%%%%%%%%%%%%%%%%%%%%%%
%%%%%%%%%%%%%%%%%%%%%%%%%%%%%%%%%%%%%%%%%%%%%%%%%%%
\subsection{Case Study 4: Moderate Attacker}
Moderate attackers possess a deeper understanding of the system than naive or amateur attackers. They can identify the limitations of previous attack strategies and adapt accordingly. Specifically, a moderate attacker knows that a single device ID cannot exceed the per-day request threshold $\mathrm{R}$ without being flagged by the server. To circumvent this restriction, the attacker employs multiple physical devices—five in our case—to interact with the original PWA and retrieve valid, encrypted URLs in real time. These URLs are stored in a centralized database accessible to all instances of the distributed malicious application. The attacker also knows the user request distribution and, based on this information, continuously generates valid URLs using the five attacker devices. 

\begin{figure}[t]
\centering
\includegraphics[width = 0.9\textwidth]{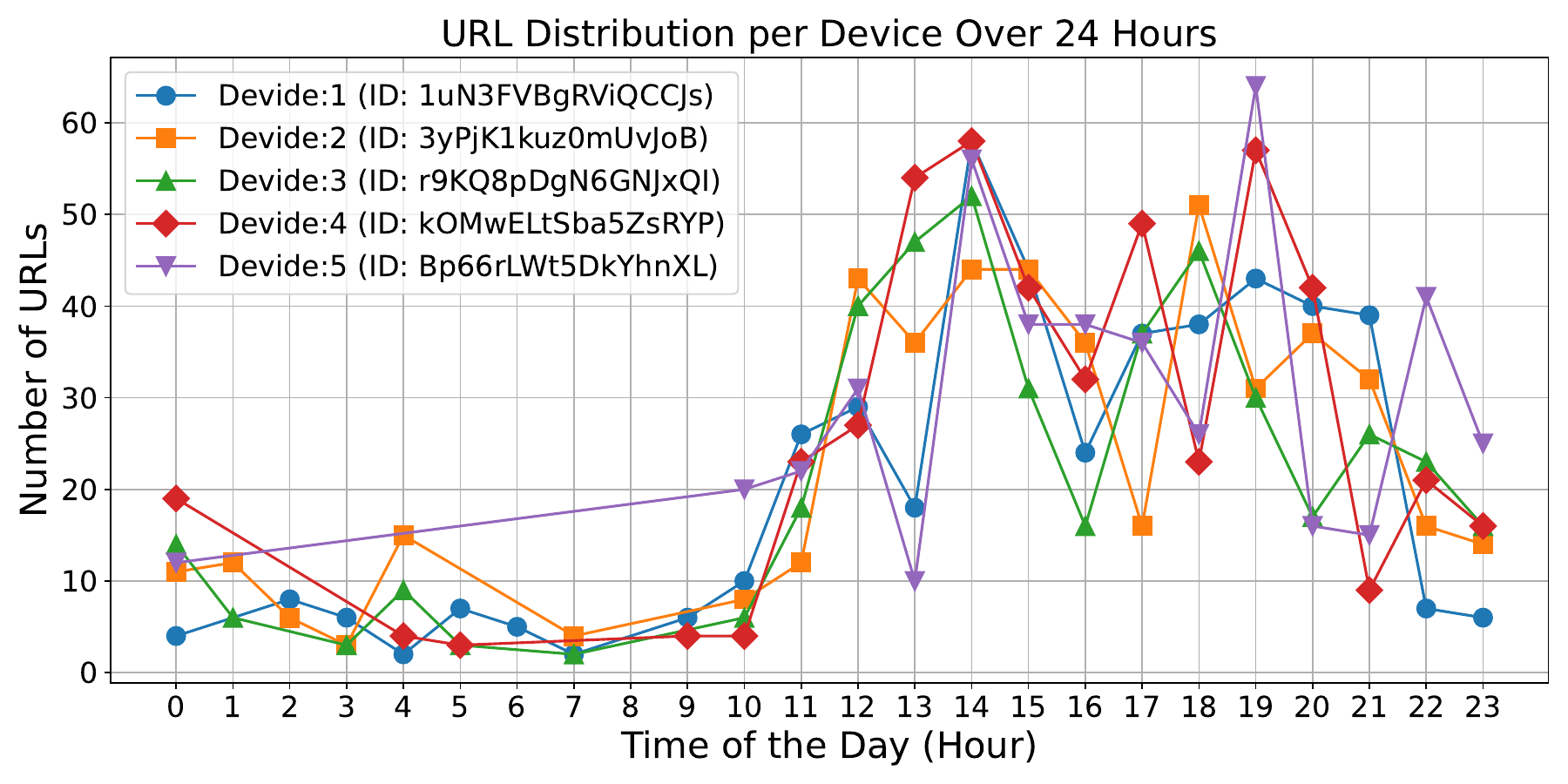}
\centering
%\vspace{-9pt}
\caption{Number of URLs generated at a certain time by an attacker device.}
\label{fig:case-study-url-distribution}
%\vspace{-15pt}
\end{figure}

Figure~\ref{fig:case-study-url-distribution} illustrates the number of URLs generated by the attacker's devices at different hours of the day. These URLs are then distributed to 20 malicious app users. However, the attacker applies a random distribution strategy, which results in the same URL (with identical timestamp and device ID) being assigned to multiple users. Consequently, the server detects these replication attempts and denies the requests. In our case study, 707 requests were made, averaging 35.35 requests per user. These were handled by the five attacker devices, which received an average of 134.80 requests each. However, as the per-device threshold was set to 30 requests, most of the requests exceeded the limit. As a result, only 150 requests (21.2\%) were successful, while the server denied the remaining 557 requests (78.8\%).

%%%%%%% URL Distribution %%%%%
%%%%%%%%%%%%%%%%%%%%%%%%%%%%%%

%%%%%%%%%%%%%%%%%%%%%%%%%%%%%%%%%%%%%%%%%%%%%%%%%%%
%%%%%%%%%%%%%%%%%%%%%%%%%%%%%%%%%%%%%%%%%%%%%%%%%%%
\subsection{Case Study 5: Sophisticated Attacker}
We now consider a more advanced attacker who employs a round-robin distribution strategy to rotate the use of valid URLs among malicious app users. This approach avoids triggering request anomalies by ensuring that no single URL or device ID exceeds the per-device request threshold. In this case, the attacker immediately deletes a URL from the centralized database once it has been distributed to a user, preventing reuse and reducing the likelihood of detection. As a result, the success rate of the attack improves compared to the previous scenario. However, given that the per-device threshold remains set at 30 requests, many still exceed the limit. Out of a total of 716 requests, 300 (41.9\%) were successfully served, while the remaining 416 requests (58.1\%) were denied by the server.
\section{Dataset Details}

To simulate realistic app usage patterns, we developed a synthetic dataset generator that models user-device interactions over a 24-hour period. The system simulates encrypted user requests, timestamps, and device identifiers, ensuring behavioral diversity and cryptographic consistency across users and time.

\noindent\underline{\textit{User and Device Modeling}} The dataset consists of $N$ users, each generating a random number of requests drawn from a uniform distribution between 25 and 45 requests. Each user is assigned a set of devices from a pool of $D$ unique device identifiers. Device assignment follows a round-robin mechanism to simulate multiple users sharing overlapping device sets, mimicking real-world shared environments or spoofed device identities.

\noindent\underline{\textit{Temporal Distribution of Requests}} To capture realistic app usage behavior, the hourly request distribution follows a normalized, empirically-inspired pattern.

%%%%%%%%%%%% Scalability %%%%%%%%%%%%%%%
\begin{figure}[!t]
%\vspace{-6pt}
    \begin{center}
         \subfigure[]
        {
        \label{fig:DatasetDistribution}
            \includegraphics[width=0.45\columnwidth]{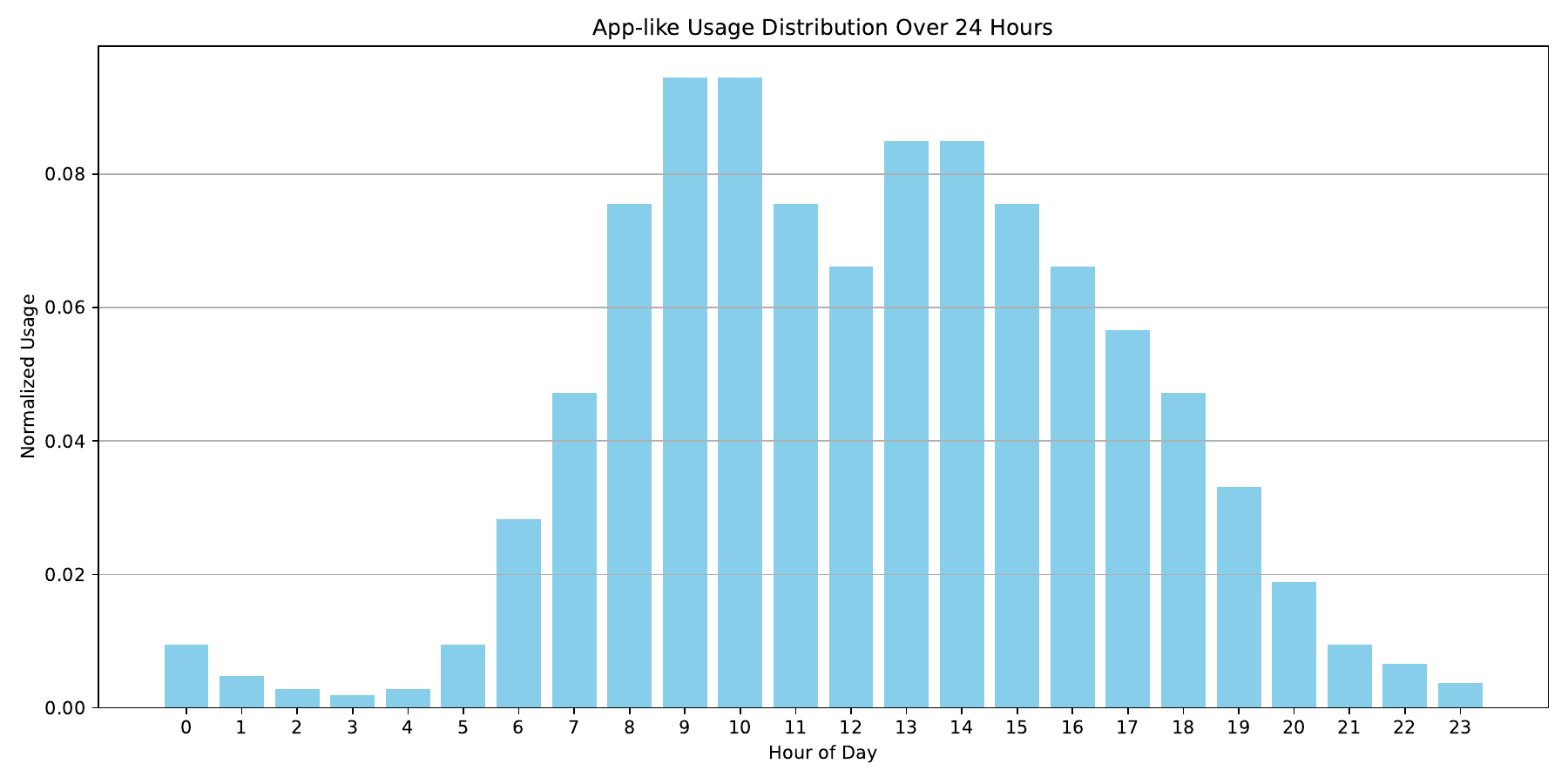}
        }
        \hspace{-8pt}
        \subfigure[]
        {
        \label{subfig:user-requests}
            \includegraphics[width=0.45\columnwidth]{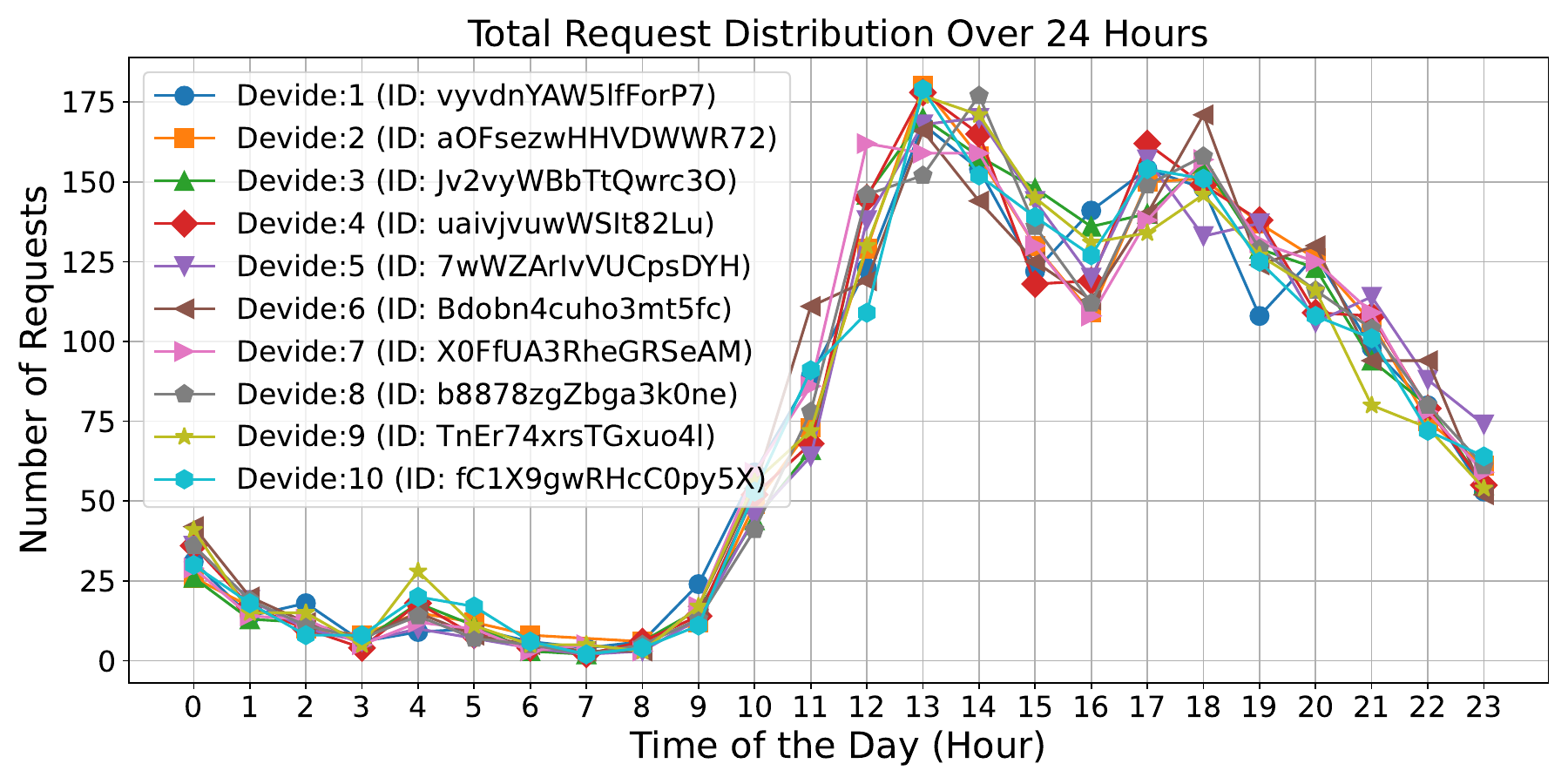}
        }
        \hspace{-8pt}
    \end{center}
    \vspace{-20pt}
    \caption{(a) User request distribution, (b) number of requests issued by users at different hours following the distribution.}
    \label{fig:data-dictribution}
    \vspace{-15pt}
\end{figure}
%%%%%%%%%%%%%%%%%%%%%%%%%%%

% \begin{figure}[!htbp]
% \centering
% \includegraphics[width=0.9\textwidth]{figures/Dataset_Distribution.pdf}
% \caption{User Request Distribution.}
% \label{fig:DatasetDistribution}
% \end{figure}

As illustrated in Figure~\ref{fig:DatasetDistribution}, activity is minimal during late-night hours (12 AM–5 AM), increases significantly in the morning, peaks between 9 AM–11 AM, and gradually declines in the evening. Request timestamps are sampled by weighting hourly selection with this usage distribution and then randomly choosing a second within the selected hour. Figure~\ref{subfig:user-requests} shows the number of requests issued by 10 devices at different hours, following the specified distribution. The maximum number of requests did not exceed 175.

\subsection{Encrypted Parameter Generation}

Each request contains a timestamp and a device identifier, concatenated and formatted as \texttt{timestamp@PWA\\deviceID}. These raw data strings are encrypted using AES in CBC mode with a predefined 16-byte secret key and initialization vector (IV). This ensures confidentiality and simulates real-world obfuscation of query parameters. Encryption and decryption validity are asserted for every entry to maintain consistency.
\subsection{Benign and Malicious User Dataset Configuration}

We generate two distinct types of datasets: one for benign users and another for malicious users, each designed to reflect different behavioral patterns in request generation and device usage.

\noindent\underline{\textit{Benign Users:}} For the benign user scenario, we synthesize datasets for three population sizes: 100, 200, and 300 users. Each benign user is assigned a unique device, ensuring a one-to-one mapping between user and device identifiers. Request generation for each user follows the app-like hourly usage distribution discussed previously (see Figure~\ref{fig:DatasetDistribution}), representing natural, human-driven app interaction patterns. This setting serves as the baseline for evaluating normal system behavior.

\noindent\underline{\textit{Malicious Users:}} In contrast, the malicious user datasets are designed to simulate adversarial behavior with potential device spoofing or identity duplication. For each user population size (100, 200, 300 users), we simulate attacks using 10, 20, and 30 unique device identifiers, respectively. These device identifiers are cyclically reused across users, simulating scenarios where attackers attempt to impersonate multiple users using limited physical devices. As with benign users, the request timestamps for malicious users are also sampled using the same temporal distribution model to ensure comparability.

\noindent\underline{\textit{Comparative Objective:}} This two-tiered generation approach enables a robust evaluation of system performance under normal (benign) and adversarial (malicious) conditions. By varying the number of users and attack devices, we can analyze the impact of user density and device reuse on system accuracy, failure rates, and detection capabilities.

\section{Evaluation}
\label{sec:evaluation}
This section presents the findings from evaluating the proposed PWA defense against an advanced attack strategy. We present the results of SPARE's evaluation by addressing the following research questions.

\textbf{RQ1} What is the impact of the device access threshold for the benign scenario? (Section~\ref{subsubsec:eval-benign})

\textbf{RQ2} What are the framework findings in assessing the proposed attack impact with variable system parameters? (Section~\ref{subsubsec:eval-system-params})

\textbf{RQ3} What are the framework findings in assessing the proposed attack impact with variable attacker's capability? Section~\ref{subsubsec:eval-attackers-capability})

% \begin{itemize}
%     \item \textit{Define attacker success rate: number of requests that he has got responses to in different scenarios}
%     \item \textit{User dissatisfaction}
%     \item \textit{Actual user using malicious application - expected time of operation}
%     \item \textit{User dissatisfaction probability - during peak hours (peak hour analysis)}
%     \item \textit{Trend analysis of user}
% \end{itemize}

%\textbf{Distributions:} uniform, random, Zipfian distribution (?)
%%%%%%%%%%%%%%%%%%%%%%%%%%%%%%%%%%%%%%%%%%%%

\subsection{System Evaluation in Benign Scenario}
\label{subsubsec:eval-benign}

\begin{table}[htbp]
\centering
\caption{Benign User Request Statistics}
\resizebox{\textwidth}{!}{%
\begin{tabular}{|c|c|c|c|c|c|}
\hline
\textbf{No of User} & \textbf{Threshold} & \textbf{Total Requests} & \textbf{Successful Requests} & \textbf{Failed Requests} & \textbf{Failed Percentage} \\
\hline
100 & 30 & 3558 & 2942 & 616 & 17.31 \\
\hline
100 & 35 & 3558 & 3263 & 295 & 8.29 \\
\hline
100 & 40 & 3558 & 3477 & 81  & 2.28 \\
\hline
200 & 30 & 6937 & 5832 & 1105 & 15.93 \\
\hline
200 & 35 & 6937 & 6441 & 496 & 7.15 \\
\hline
200 & 40 & 6937 & 6796 & 141 & 2.03 \\
\hline
300 & 30 & 10347 & 8803 & 1544 & 14.92 \\
\hline
300 & 35 & 10347 & 9577 & 770 & 7.44 \\
\hline
300 & 40 & 10347 & 10013 & 334 & 3.23 \\
\hline
\end{tabular}%
}
\label{tab:benign_users_nodevices}
\end{table}
\vspace{-10pt}
\subsubsection{Benign User Performance Analysis}

System performance was rigorously evaluated under varying benign user population sizes and corresponding request thresholds. The experimental dataset encompassed configurations for 100, 200, and 300 users, each simulated with a unique device identifier, coupled with request thresholds set at 30, 35, and 40.

\begin{wrapfigure}{r}{0.5\columnwidth}
\includegraphics[width=0.5\textwidth]{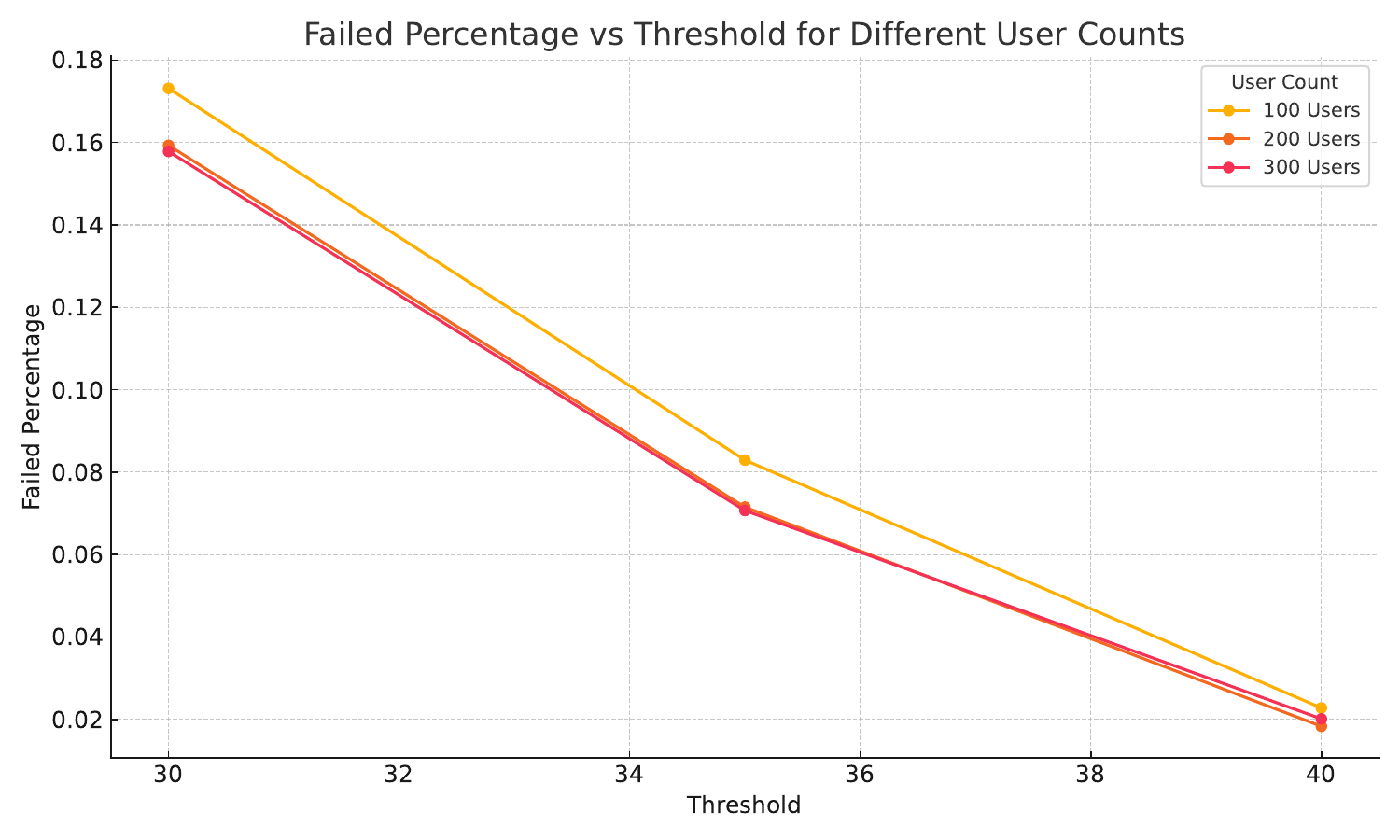}
\caption{Failed percentage vs threshold for different user counts.}
\label{fig:failed_percentage_plot}
\vspace{-25pt}
\end{wrapfigure}

Figure~\ref{fig:failed_percentage_plot} graphically presents the relationship between the failure percentage and the applied threshold across these distinct user counts. Consistent with expectations, an increase in the request threshold demonstrably leads to a significant reduction in the observed failure rate. For example, within the 100-user configuration, the failure percentage decreased from 17.31\% at a threshold of 30 to a mere 2.28\% when the threshold was elevated to 40. A comparable trend was observed for the 200 and 300-user scenarios, although the absolute failure rates were marginally higher, which can be attributed to the increased system load associated with larger user populations.

%%%%%%%%%%%%%%%%%%%%%%%%%%%%%%%%%%%%%%%%%%%%%%
%%%%%%%%%%%%%%%%%%%%%%%%%%%%%%%%%%%%%%%%%%%%%%

This observed trend underscores the critical importance of judicious threshold tuning in preserving system reliability. Lower thresholds, while potentially enhancing sensitivity to anomalous behavior, are inherently more restrictive, which can result in elevated rejection rates for benign traffic. Conversely, higher thresholds facilitate the passage of a greater volume of benign traffic, thereby improving user experience, but may inadvertently diminish the system's ability to detect genuine anomalies. Consequently, the selection of an optimal threshold necessitates a careful equilibrium between detection accuracy and seamless user experience.

% \begin{figure}[!htbp]
% \centering
% \includegraphics[width=0.8\textwidth]{figures/benign_failed_percentage_vs_threshold.pdf}
% \caption{Failed Percentage vs Threshold for Different User Counts}
% \label{fig:failed_percentage_plot}
% \end{figure}

%%%%%%%%%%%%%%%%%%%%%%%%%%%%%%%%%%%%%%%%%%%%
\subsection{Attack Impact Evaluation by Varying System Parameters}
\label{subsubsec:eval-system-params}

\subsubsection{Malicious User Scenario Analysis}

\begin{table}[!htbp]
\centering
\caption{Malicious User Request Statistics (Failure Rates in Percentage)}
\resizebox{\textwidth}{!}{%
\begin{tabular}{|c|c|c|c|c|c|c|}
\hline
\textbf{No of User} & \textbf{Attack Devices} & \textbf{Threshold} & \textbf{Total Requests} & \textbf{Successful Requests} & \textbf{Failed Requests} & \textbf{Failed Percentage} \\
\hline
100 & 10 & 30 & 3524 & 300 & 3224 & 91.49\% \\
\hline
100 & 20 & 30 & 3580 & 600 & 2980 & 83.24\% \\
\hline
100 & 30 & 30 & 3466 & 870 & 2596 & 74.90\% \\
\hline
100 & 10 & 35 & 3524 & 350 & 3174 & 90.07\% \\
\hline
100 & 20 & 35 & 3580 & 700 & 2880 & 80.45\% \\
\hline
100 & 30 & 35 & 3466 & 1013 & 2453 & 70.77\% \\
\hline
100 & 10 & 40 & 3524 & 400 & 3124 & 88.65\% \\
\hline
100 & 20 & 40 & 3580 & 798 & 2782 & 77.71\% \\
\hline
100 & 30 & 40 & 3466 & 1152 & 2314 & 66.76\% \\
\hline
200 & 10 & 30 & 6922 & 300 & 6622 & 95.67\% \\
\hline
200 & 20 & 30 & 6859 & 600 & 6259 & 91.25\% \\
\hline
200 & 30 & 30 & 7040 & 900 & 6140 & 87.22\% \\
\hline
200 & 10 & 35 & 6922 & 350 & 6572 & 94.94\% \\
\hline
200 & 20 & 35 & 6859 & 700 & 6159 & 89.79\% \\
\hline
200 & 30 & 35 & 7040 & 1050 & 5990 & 85.09\% \\
\hline
200 & 10 & 40 & 6922 & 400 & 6522 & 94.22\% \\
\hline
200 & 20 & 40 & 6859 & 800 & 6059 & 88.34\% \\
\hline
200 & 30 & 40 & 7040 & 1197 & 5843 & 83.00\% \\
\hline
300 & 10 & 30 & 10413 & 300 & 10113 & 97.12\% \\
\hline
300 & 20 & 30 & 10388 & 600 & 9788 & 94.22\% \\
\hline
300 & 30 & 30 & 10331 & 900 & 9431 & 91.29\% \\
\hline
300 & 10 & 35 & 10413 & 350 & 10063 & 96.64\% \\
\hline
300 & 20 & 35 & 10388 & 700 & 9688 & 93.26\% \\
\hline
300 & 30 & 35 & 10331 & 1050 & 9281 & 89.84\% \\
\hline
300 & 10 & 40 & 10413 & 400 & 10013 & 96.16\% \\
\hline
300 & 20 & 40 & 10388 & 800 & 9588 & 92.30\% \\
\hline
300 & 30 & 40 & 10331 & 1200 & 9131 & 88.38\% \\
\hline
\end{tabular}%
}
\label{tab:malicious_users}
\end{table}
Table~\ref{tab:malicious_users} comprehensively details the system's performance under various malicious conditions, examining 27 distinct configurations. The experimental design systematically manipulated three key parameters: the number of active users (100, 200, 300), the quantity of attack devices (10, 20, 30), and the predetermined rejection threshold (30, 35, 40). Each row within the table represents a unique test scenario, meticulously simulating coordinated malicious behavior with varying levels of device distribution and system sensitivity to threats.

Consistently across all tested configurations, the failure percentage, which quantifies the proportion of rejected requests, remained notably elevated. This persistently high rejection rate provides substantial validation for the system's robust capability to accurately identify and effectively block suspicious network traffic. The findings underscore the system's efficacy in maintaining operational integrity amidst diverse and escalating malicious activities.

These results demonstrate the system's resilience and its proficiency in threat mitigation, highlighting its potential for deployment in environments requiring stringent security protocols. The systematic variation of parameters allows for a nuanced understanding of how changes in user load, attack vector density, and detection sensitivity influence overall system robustness against adversarial actions. Several salient patterns emerged from the analysis of system performance under malicious conditions:

\noindent\underline{\textit{Impact of Threshold:}} An increase in the rejection threshold from 30 to 40 consistently correlated with a reduction in the observed failure percentage. For instance, in the configuration with 300 users and 10 attack devices, the failure rate decreased from 97.12\% at a threshold of 30 to 96.16\% at a threshold of 40. This suggests that elevated thresholds confer a marginal increase in system permissiveness, allowing a slightly greater proportion of requests to traverse the system.

\noindent\underline{\textit{Impact of Attack Device Count:}} For a constant user population and a fixed rejection threshold, an increase in the number of attack devices corresponded to a lower failure percentage. As an illustration, with a threshold of 35 and 100 users, the failure rate decreased from 90.07\% (10 attack devices) to 70.77\% (30 attack devices). This phenomenon indicates that attacks dispersed across a larger number of devices are inherently more challenging to detect, as the pattern of device reuse, a potential indicator of malicious activity, becomes less pronounced.

\noindent\underline{\textit{Impact of User Volume:}} While the total request volume scaled proportionally with an increase in the number of users, the system consistently maintained robust detection performance. At a threshold of 30 with 10 attack devices, the failure rates were observed as 91.49\%, 95.67\%, and 97.12\% for 100, 200, and 300 users, respectively. This demonstrates the system's strong scalability in effectively rejecting anomalous traffic even under increased overall load.

This extensive simulation comprehensively highlights the critical imperative of balancing detection aggressiveness, primarily controlled via threshold settings, against the evolving nature of adversarial strategies, such as the distribution of attack devices. Furthermore, the findings unequivocally validate the system's capacity to maintain a high rate of malicious traffic rejection, even when confronted with escalating attack volumes.

%%%%%%%%%%%%%%%%%%%%%%%%%%%%%%%%%%%%%%%%%%%%%%%%
% \begin{figure}[htbp]
% \centering
% \includegraphics[width=0.8\textwidth]{figures/3D_Surface_Failed_Percentage.pdf}
% \caption{3D surface plot showing failed percentage as a function of number of users and attack devices at a fixed threshold (Threshold = 30).}
% \label{fig:3d_surface}
% \end{figure}
%%%%%%%%%%%%%%%%%%%%%%%%%%%%%%%%%%%%%%%%%%%%%%%%%%%%%%%%%%
%%%%%%%%%%%%%%%%%%%%%%%%%%%%%%%%%%%%%%%%%%%%%%%%%%%%%%%%%%

\subsubsection{3D Surface Analysis of Failure Behavior}
\begin{wrapfigure}{r}{0.5\columnwidth}
\includegraphics[width=0.5\textwidth]{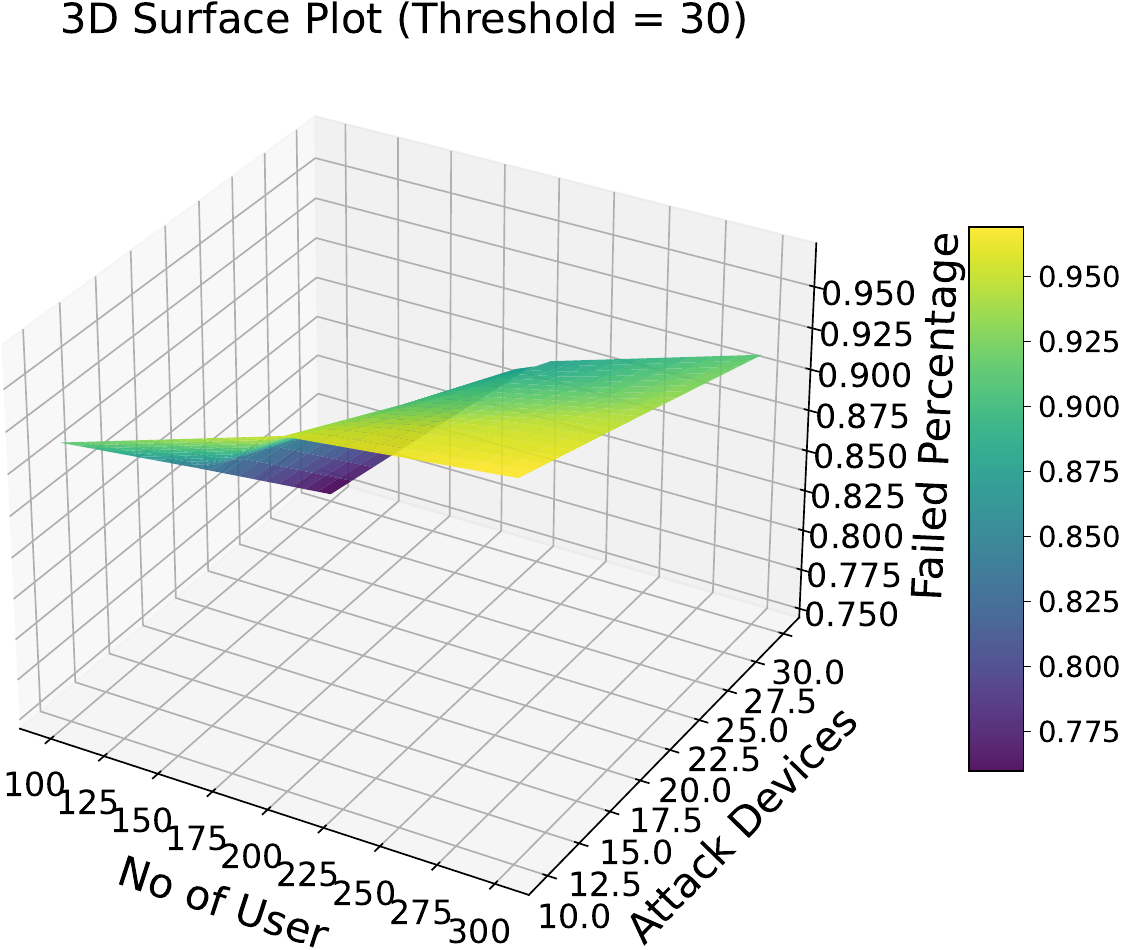}
\caption{3D surface plot showing failed percentage as a function of number of users and attack devices at a fixed threshold (Threshold = 30).}
\label{fig:3d_surface}
\vspace{-15pt}
\end{wrapfigure}

Figure~\ref{fig:3d_surface} illustrates a three-dimensional surface plot depicting the failed percentage as a function of both the number of users and the count of attack devices, with the detection threshold held constant at 30. This visualization offers a comprehensive perspective on system behavior under varying operational load and adversarial distributions.

The surface plot reveals two primary trends:

\noindent\underline{\textit{Increase in users leads to higher failure rates:}} An increase in the number of users corresponds to an elevated failure rate, manifested as an ascent along the failure percentage axis. This observation is consistent with the principle that a larger user base generates a higher volume of requests, consequently augmenting system workload and increasing the probability of rejection when device reuse patterns are pronounced.
    
\noindent\underline{\textit{Increase in attack devices reduces failure percentage:}} Conversely, an increase in the number of attack devices leads to a reduction in the failure percentage, indicated by a flattening of the surface. This reflects the inherent challenge in detecting distributed attacks, wherein malicious activities are spread across numerous devices, thereby diminishing the per-device detection signal.

The curvature of the surface plot effectively demonstrates the nonlinear interaction between user volume and device diversity. When a limited number of devices are utilized in conjunction with a high user count, the failure percentage exhibits a sharp peak. However, as the number of devices increases, even substantial user loads result in diminished rejection rates. This three-dimensional perspective complements prior two-dimensional heatmaps and reinforces previous findings: centralized attacks are more readily detectable, whereas distributed attack strategies necessitate the deployment of more sophisticated countermeasures.

%%%%%%%%%%%%%%%%%%%%%%%%%%% heatmap %%%%%%%%%%%%%%%%%%%%%%%%%%%
\begin{figure}[!t]
%\vspace{-6pt}
    \begin{center}
         \subfigure[]
        {
        \label{subfig:scalability_time_horizon}
            \includegraphics[width=0.32\columnwidth]{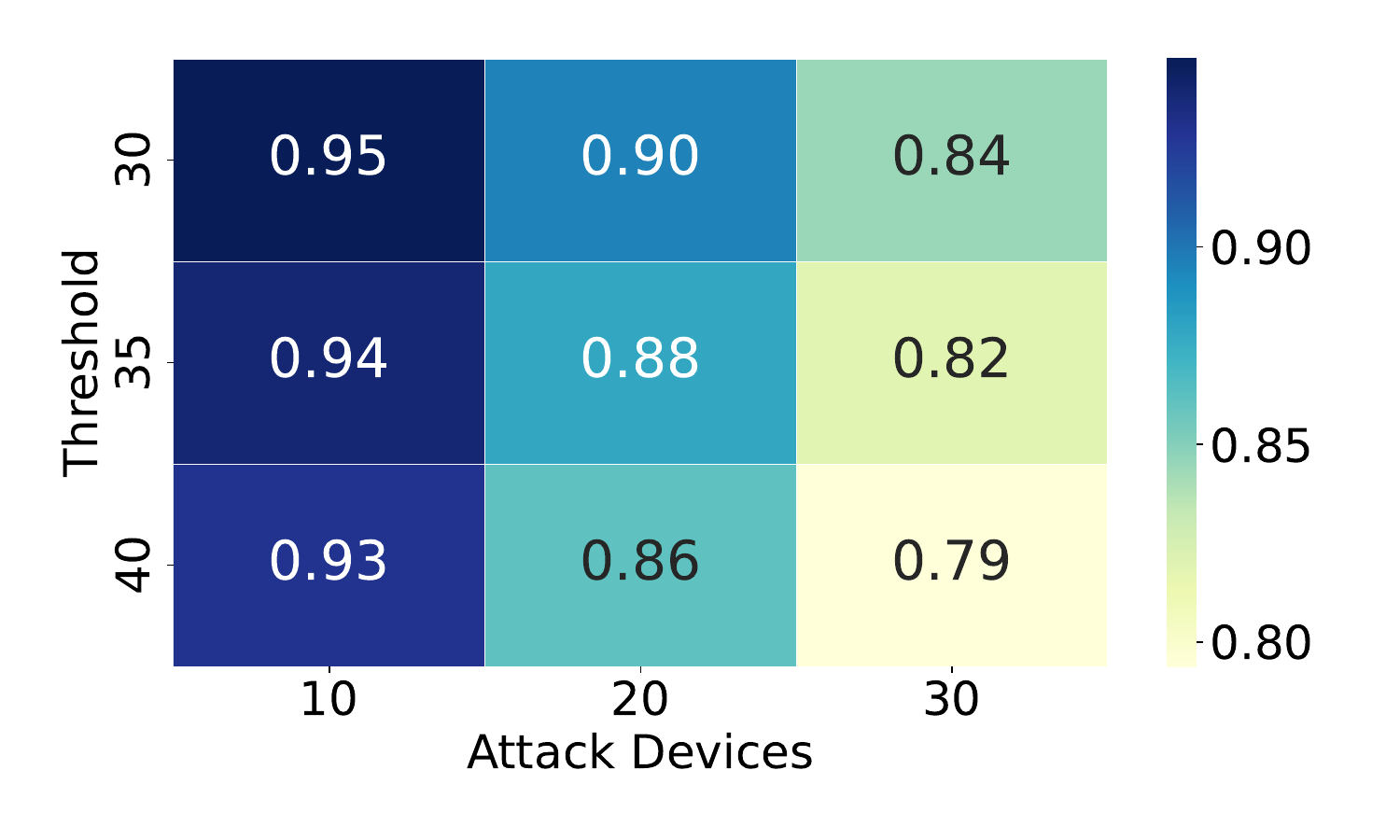}
        }
        \hspace{-8pt}
        \subfigure[]
        {
        \label{subfig:scalability_horizontal}
            \includegraphics[width=0.32\columnwidth]{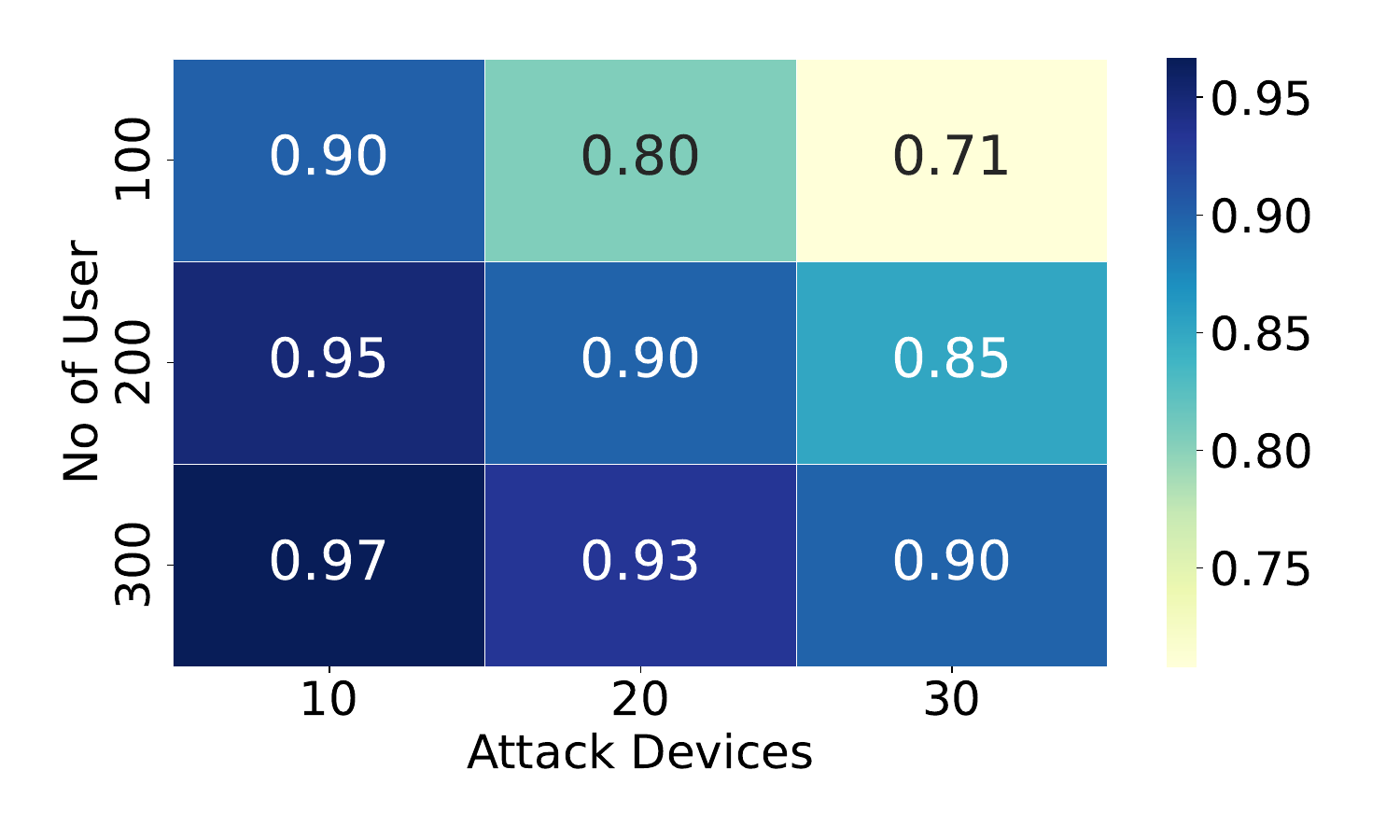}
        }
        \hspace{-8pt}
        \subfigure[]
        {
        \label{subfig:scalability_threshold}
            \includegraphics[width=0.32\columnwidth]{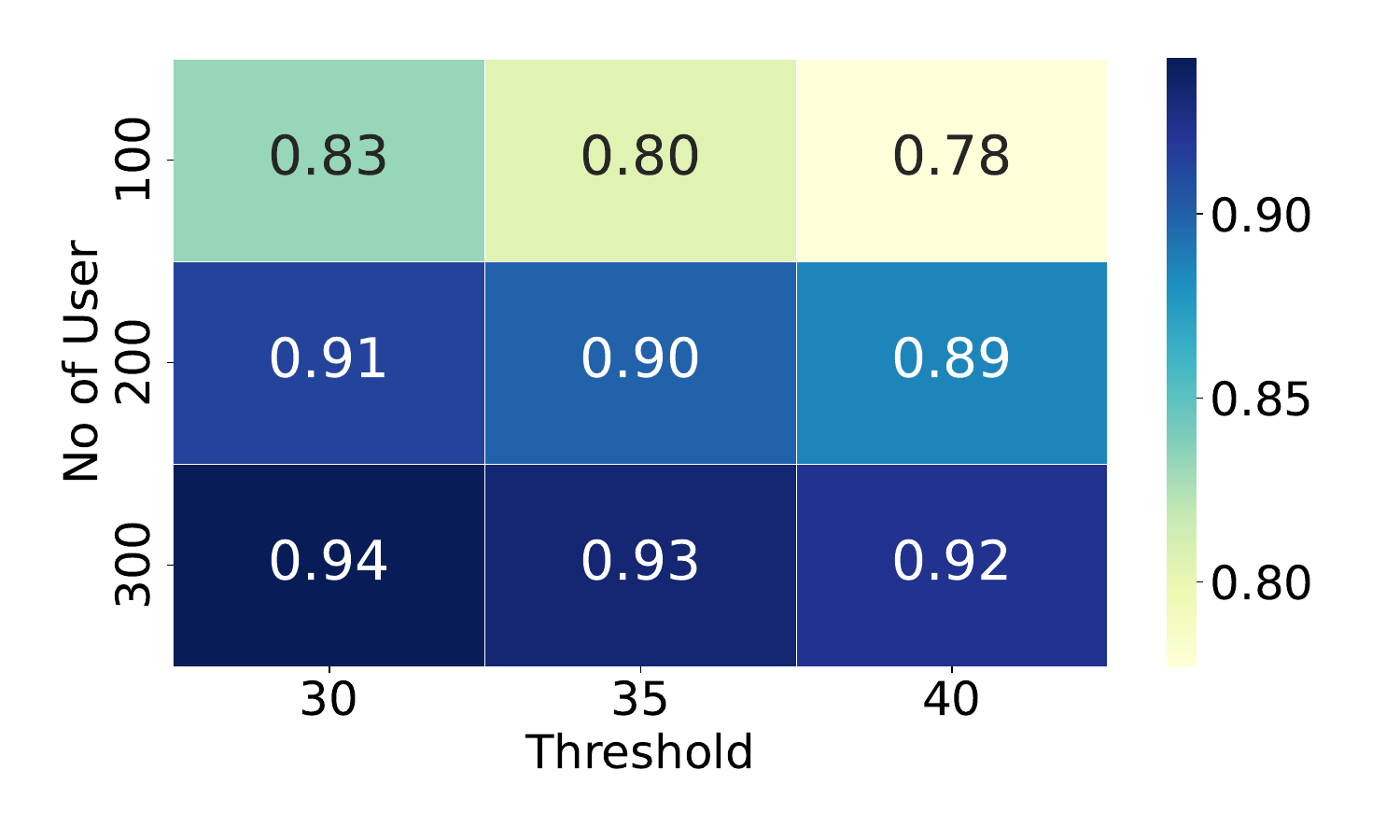}
        }
    \end{center}
    \vspace{-20pt}
    \caption{Heatmaps showing failed percentage across various parameter configurations: (a) threshold vs attack devices, (b) number of users vs attack devices, and (c) number of users vs threshold.}
    \label{fig:combined_heatmaps}
    \vspace{-15pt}
\end{figure}
%%%%%%%%%%%%%%%%%%%%%%%%%%%
This 3D perspective complements the 2D heatmaps and reinforces earlier observations: centralized attacks are easier to detect, while distributed strategies demand more sophisticated countermeasures.

%%%%%%%%%%%%%%%%%%%%%%%%%%%%%%%%%%%%%%%%%%%%%%
%%%%%%%%%%%%%%%%%%%%%%%%%%%%%%%%%%%%%%%%%%%%%%

% \begin{figure}[htbp]
% \centering

% \textbf{(a)}~Threshold vs Attack Devices\\
% \includegraphics[width=0.3\textwidth]{figures/heatmap_threshold_vs_devices.pdf}

% \vspace{0.6cm}

% \textbf{(b)}~No of Users vs Attack Devices\\
% \includegraphics[width=0.3\textwidth]{figures/heatmap_users_vs_devices.pdf}

% \vspace{0.6cm}

% \textbf{(c)}~No of Users vs Threshold\\
% \includegraphics[width=0.3\textwidth]{figures/heatmap_users_vs_threshold.pdf}

% \caption{Heatmaps showing failed percentage across various parameter configurations:
% (a) threshold vs attack devices,
% (b) number of users vs attack devices,
% and (c) number of users vs threshold.}
% \label{fig:combined_heatmaps}
% \end{figure}

\subsubsection{Heatmap-Based Analysis of Failure Patterns}

Figure~\ref{fig:combined_heatmaps} presents a series of heatmaps that delineate the variation in failed request percentage across three distinct parameter dimensions. Each subfigure illustrates how failure behavior changes as a function of the interaction between two parameters, with the third parameter held constant.

\noindent\underline{\textit{(a) Threshold vs Attack Devices:}} Subfigure (a) examines the relationship between the number of attack devices and the system's detection threshold. A decrease in the failure percentage is observed as the number of attack devices increases. This is anticipated, as distributing malicious activity across a greater number of devices reduces the likelihood of detection due to diminished device reuse. Conversely, an elevation of the threshold renders the system more permissive, consequently reducing failure rates. Notably, the confluence of a low threshold and a limited number of attack devices yields the highest failure rates, underscoring the system's efficacy under stringent security conditions and low device diversity.

\noindent\underline{\textit{Number of Users vs Attack Devices:}}  Subfigure (b) investigates the impact of user count and attack device count on the failure percentage. As the number of users increases, the system processes a greater volume of requests, yet it maintains high rejection rates when the number of attack devices is small. For instance, a scenario involving 300 users with only 10 attack devices exhibits a very high failure percentage, attributable to dense device reuse. As the number of attack devices increases, failure rates decrease across all user sizes, emphasizing the inherent difficulty in detecting distributed attacks where each device simulates fewer users. 

\noindent\underline{\textit{(c) Number of Users vs Threshold:}} Subfigure (c) analyzes the system's sensitivity to varying threshold values across different user counts. For all user sizes, increasing the threshold consistently results in a lower failure percentage, illustrating the trade-off between stringency and permissiveness. Nevertheless, even at elevated thresholds, the system maintains reasonably high detection capabilities when the number of users is substantial, demonstrating its scalability and robustness.

\subsection{Predictive Modeling of Failed Requests}

To estimate the failed percentage of malicious users based on system parameters, we developed regression models using real simulation data. The models consider three variables: number of users (\textit{U}), number of attack devices (\textit{D}), and request threshold (\textit{T}).

\begin{figure}[H]
\centering
\includegraphics[height=0.3\textheight]{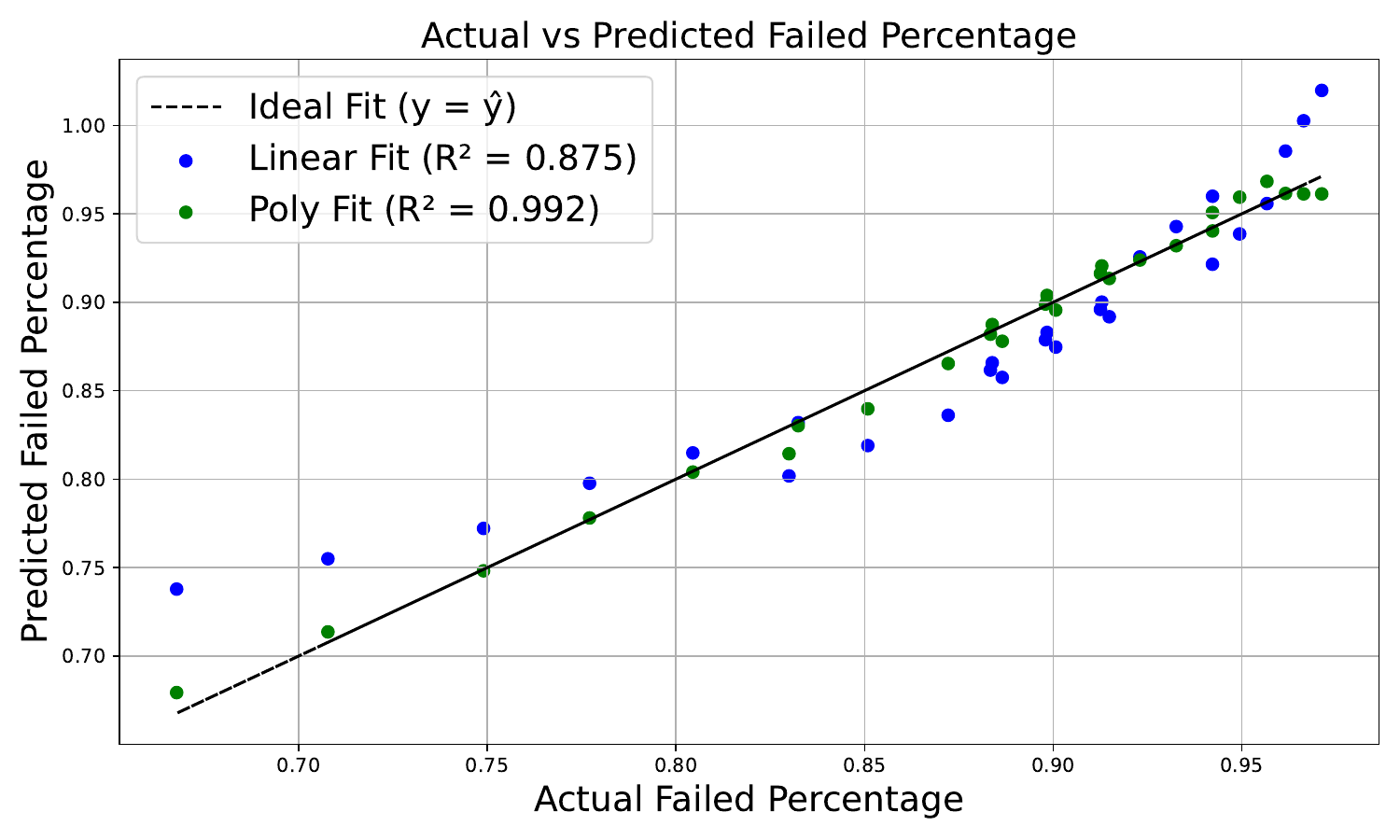}
\caption{Actual vs Predicted Failed Percentage for Malicious Users (Linear and Polynomial Models)}
\label{fig:regression_fit}
\end{figure}

\paragraph{Linear Model.}  
The linear regression equation is:

\[
\text{Failed\%} = 0.9906 + 0.0006 \cdot U - 0.0060 \cdot D - 0.0034 \cdot T
\]

This model explains 87.5\% of the variance (\( R^2 = 0.875 \)), indicating a strong fit. However, it assumes a linear relationship between input variables and the failed percentage.

\paragraph{Polynomial Model.}  
To capture more complex interactions, we also fitted a second-degree polynomial regression model. The equation includes squared and interaction terms:

\[
\begin{aligned}
\text{Failed\%} =\ & 1.0277 + 0.0006 \cdot U - 0.0066 \cdot D - 0.0040 \cdot T \\
& - 3.1 \times 10^{-6} \cdot U^2 + 3.1 \times 10^{-5} \cdot (U \cdot D) + 1.8 \times 10^{-5} \cdot (U \cdot T) \\
& + 5.9 \times 10^{-6} \cdot D^2 - 1.7 \times 10^{-4} \cdot (D \cdot T) + 4.8 \times 10^{-6} \cdot T^2
\end{aligned}
\]

This model achieved a higher \( R^2 = 0.992 \), indicating that the polynomial model better captures the underlying patterns in the data.

Figure~\ref{fig:regression_fit} shows the comparison of actual vs. predicted failed percentages for malicious users, showing that the polynomial model outperforms the linear model in capturing nonlinear patterns in user behavior.

%%%%%%%%%%%%%%%%%%%%%%%%%%%%%%%%%%%%%%%%%%%%
%\subsection{Attack Impact Evaluation by Varying Attacker's Capability}
\label{subsubsec:eval-attackers-capability}
\subsection{Comparison Between Malicious and Benign Users with the same number of requests}
\label{subsubsec:compare-benign-malicious}

\subsubsection{Impact of Defense Technique on First Error Detection}
\begin{figure}[htbp]
\centering
\includegraphics[width=0.85\textwidth]{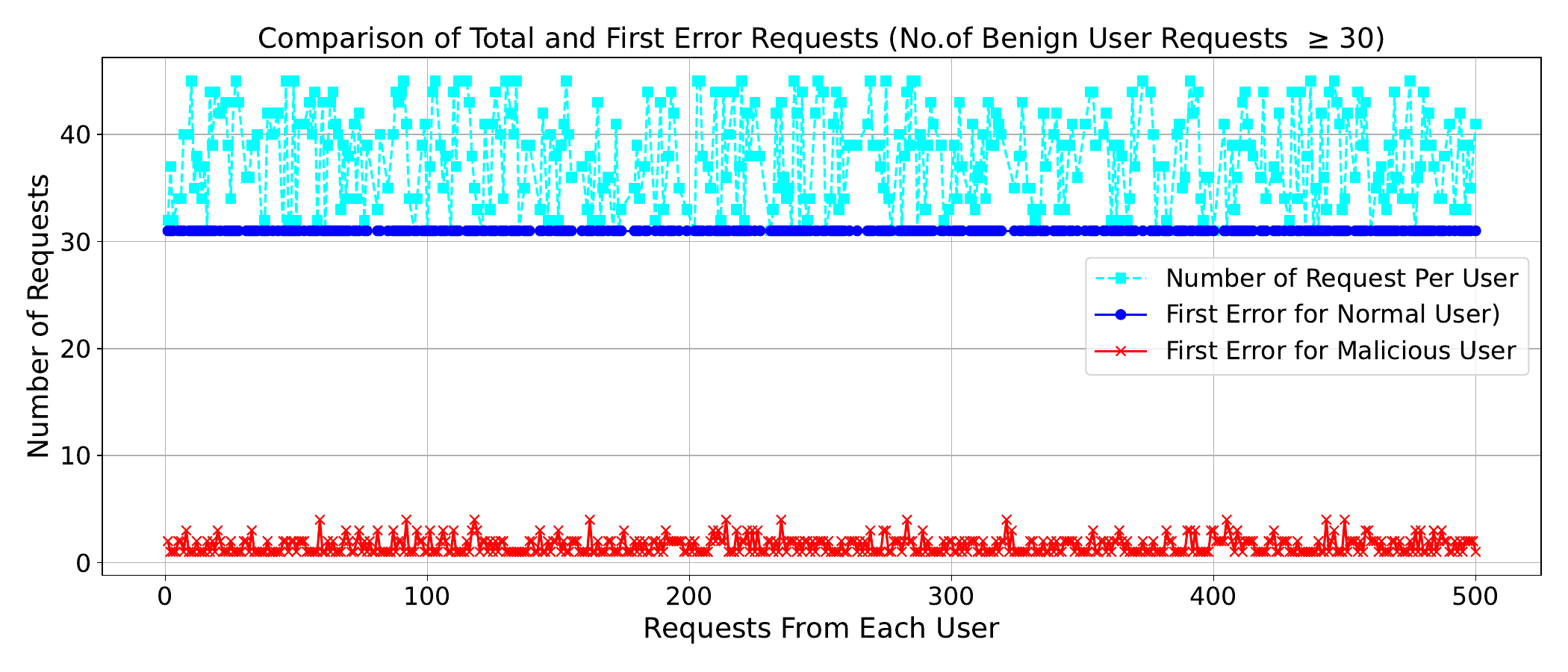}
\caption{Comparison of total requests and first error occurrences for benign and malicious users (threshold = 30).}
\label{fig:first_error_comparison}
\end{figure}

Figure~\ref{fig:first_error_comparison} provides a comparative analysis of benign and malicious users, based on their total request count and the precise position of their initial rejected request, under a consistent system threshold of 30 requests per user.

% \begin{figure}[htbp]
% \centering
% \includegraphics[width=0.85\textwidth]{figures/3D_scatter_malicious_vs_normal.pdf}
% \caption{3D scatter plot comparing failure patterns in malicious and normal scenarios based on number of users, attack devices, and threshold settings.}
% \label{fig:scatter_malicious_normal}
% \end{figure}

Figure~\ref{fig:scatter_malicious_normal} presents a three-dimensional scatter plot that visually delineates the separation between malicious and normal (benign) request scenarios across three key operational parameters: the number of users, the count of attack devices, and the system threshold. 
% The distribution of data points within this plot reveals distinct patterns for each user group.

% \begin{itemize}
%     \item \textbf{Normal users} are typically clustered at higher threshold values and exhibit reduced variability across the dimension representing the number of attack devices. This clustering is consistent with expected, rule-abiding usage patterns, indicating predictable system interaction.

%     \item \textbf{Malicious users}, conversely, are dispersed across a broader spectrum of attack device counts and are frequently associated with lower threshold values. Their spatial dispersion within the plot effectively reflects the inherent unpredictability and aggressive nature characteristic of adversarial behavior.
%     %
%     \item The \textit{clear spatial separation} observed between malicious and normal data points in this three-dimensional space strongly suggests the viability of employing multivariate analysis or machine learning classification techniques to differentiate between these two distinct user groups robustly.
% \end{itemize}
% %%%%%%%%%%%%%%%%%%%%%%%%%%%%%%%%%%%%%%%%%%%%%%

This visualization substantively corroborates earlier analytical findings, demonstrating that malicious activity not only deviates in terms of failure rates and initial error positions but also manifests in recognizable patterns across broader system-wide parameters. Such evidence provides a robust foundation for the future development of intelligent anomaly detection systems leveraging multi-dimensional data. A pivotal characteristic of the developed PWA is the integration of a bespoke defense mechanism, engineered for the proactive detection and mitigation of malicious behaviors prior to their culmination at a predefined threshold. Empirical data substantiate the efficacy of this methodological approach:

%%%%%%%%%%%%%%%%%%%%%%%%%%%%%%%%%%%%%%
\begin{wrapfigure}{r}{0.4\columnwidth}
\includegraphics[width=0.4\textwidth]{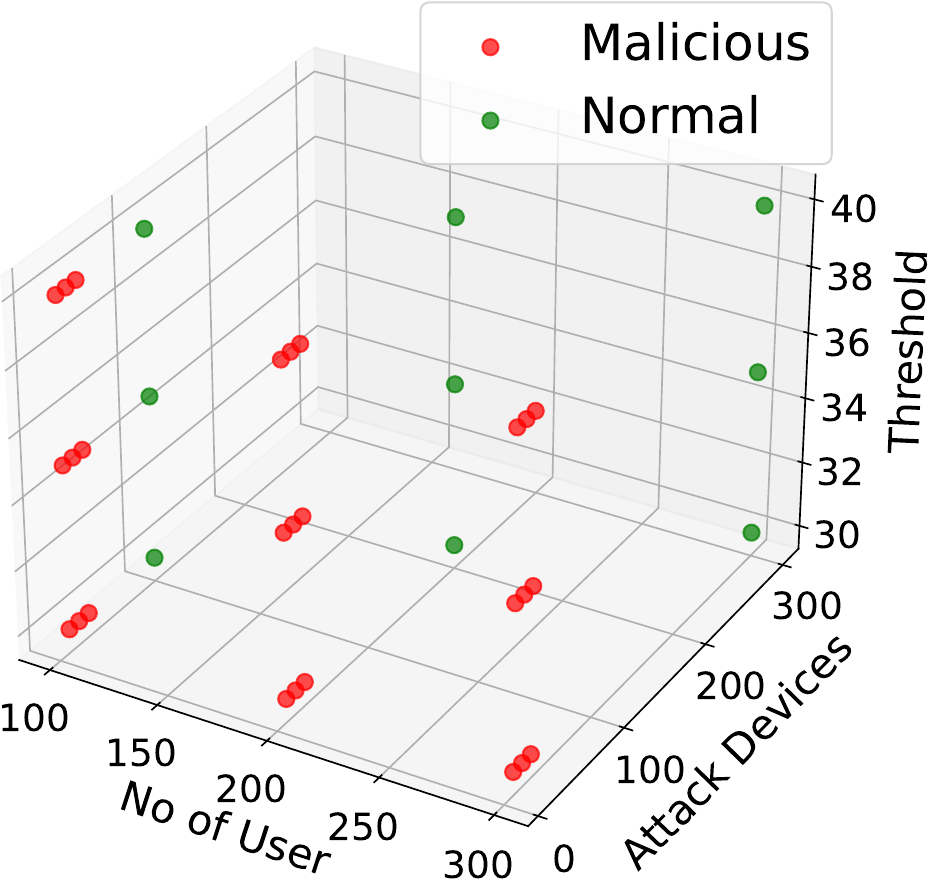}
\caption{3D scatter plot comparing failure patterns in malicious and normal scenarios based on several users, attack devices, and threshold settings.}
\label{fig:scatter_malicious_normal}
\vspace{-15pt}
\end{wrapfigure}

\noindent\underline{\textit{Benign Users:}} Benign users, whose operational profiles align with established normal behavioral parameters, typically encounter their initial error precisely at the predefined threshold limit. Their request patterns exhibit consistency, and the system permits all legitimate interactions until the application of the static threshold policy.

\noindent\underline{\textit{Malicious Users:}} In contrast, malicious users experience rejections significantly earlier in their request sequences, frequently within the initial few requests, well in advance of reaching the designated threshold of 30 requests. This observed behavior is a direct consequence of the implemented defense strategy. 

\noindent\underline{\textit{Effectiveness of the Defense:}} The premature rejection of malicious requests unequivocally demonstrates the proactive capability of the defense mechanism. Rather than solely relying on volume-based limitations, the PWA incorporates a sophisticated analysis of behavioral and contextual indicators to trigger defensive actions. This architectural choice enables the system to prevent abusive activities at a nascent stage, thereby minimizing potential adverse impacts while simultaneously preserving optimal usability for legitimate users.

These findings collectively affirm that the defense-aware PWA can effectively differentiate between benign and adversarial request patterns, culminating in the timely detection and effective mitigation of malicious activities.
\section{Discussion and Future Work}

The proposed PWA defense mechanism's performance is thoroughly evaluated by comparing the behaviors of benign and malicious users across two crucial metrics: the failed percentage and the position of the first error. This analysis elucidates the system's efficacy in distinguishing between legitimate and adversarial interactions, while also identifying critical areas for future refinement to enhance user experience and adaptive capabilities.

The \textit{Failed Percentage Analysis} consistently demonstrates the defense mechanism's robust ability to reject undesirable requests. In benign scenarios, the failure percentages remain low and are demonstrably coupled with the predefined system threshold. As the threshold is increased, the rate of benign request rejections decreases, affirming that legitimate user activities are rarely misclassified. Conversely, malicious users exhibit significantly higher failure percentages, ranging from 67\% to over 95\% across diverse configurations, including elevated thresholds and increased attack device distributions. This underscores the defense system's resilience against adversarial behavior, even under conditions of high load and distributed attacks.

Furthermore, the \textit{First Error Behavior} as depicted in Figure~\ref{fig:first_error_comparison}, illustrates another layer of differentiation in user interactions. Benign users typically encounter their initial rejections at or marginally beyond the established threshold (e.g., after approximately 30 requests), signifying consistent and acceptable user patterns. In stark contrast, malicious users experience their first rejections considerably earlier in their request sequences, often within the initial few interactions. This early detection of malicious activity is a direct consequence of the implemented defense technique, which transcends mere count-based thresholds by incorporating behavioral patterns, \textit{deviceID} reuse, and timing anomalies to proactively identify suspicious users. Consequently, numerous malicious users are rejected well before reaching the nominal threshold, thereby allowing benign users to operate with minimal impedance.

Despite its efficacy, the current system's reliance on a manually set request threshold (e.g., 30 requests) presents a limitation, as this static configuration may not be optimal across varied applications or usage patterns. Future work will explore the integration of machine learning techniques to dynamically determine appropriate thresholds based on real-time user behavior and historical request data, enabling the system to adaptively respond to diverse operational contexts. Additionally, the investigation of alternative secure communication techniques for PWAs, beyond the current encrypted request model, is warranted. Robust mechanisms such as token-based authentication, device attestation, or end-to-end encryption could further fortify the system's defenses against spoofing and unauthorized access.

While the present defense strategy effectively mitigates malicious behavior by enforcing a fixed threshold per \texttt{deviceId}, it inadvertently introduces potential usability concerns for benign users. Legitimate users with high usage demands may prematurely reach the threshold and experience unwarranted request rejections, even when their activity is entirely benign. This can lead to user dissatisfaction, particularly in scenarios involving power users, enterprise applications, or systems with inherently high interaction volumes. This limitation highlights the imperative for adaptive thresholding mechanisms capable of discerning between legitimate high-frequency usage and genuine attack patterns. Static thresholds, although simple and effective for defense, can negatively impact the user experience if not meticulously calibrated to reflect anticipated user behavior.

\section{Conclusion}
\label{sec:conclusion}
The combined analysis of failed percentage and first error positions conclusively affirms the efficacy of the defense mechanism integrated within our PWA. This dual evaluation demonstrates a clear differentiation in system response between legitimate and malicious user behaviors. Benign users consistently experience uninterrupted interactions, encountering errors only upon reaching the predefined policy limits, thereby ensuring a high degree of usability. Conversely, malicious users are accurately and promptly identified based on early deviations in their behavioral patterns. These findings unequivocally validate two critical outcomes of the implemented defense strategy. Firstly, the system maintains robust usability for legitimate users, characterized by a minimal rate of false positives. Secondly, the defense technique effectively detects and disrupts malicious activity at an early stage, significantly enhancing the system's overall resilience against sophisticated attack vectors. Such a demonstrated dual capability—successfully balancing an optimized user experience with proactive security measures—underscores the indispensable role of intelligent request monitoring, extending beyond mere static thresholds. This advanced approach is vital for effectively defending modern web applications against contemporary and evolving attack strategies.

\bibliography{References}

% \section*{Acknowledgement}

% This research was partially supported by National Science Foundation (NSF) under Award \#1929183.

%\input{biography}

\end{document}